\documentclass[floats, prd, eqnum, showpacs, nofootinbib, twocolumn,eqsecnum]{revtex4-1}

\usepackage{color,graphicx}
\usepackage{amsfonts}
\usepackage{amssymb}
\usepackage{comment}
\begin{document}

\title{Gravitational lensing by two photon spheres in a black-bounce spacetime in strong deflection limits}
\author{Naoki Tsukamoto${}^{1}$}\email{tsukamoto@rikkyo.ac.jp}

\affiliation{
${}^{1}$Department of General Science and Education, National Institute of Technology, Hachinohe College, Aomori 039-1192, Japan \\
}

\begin{abstract}
We investigate gravitational lensing by a primary photon sphere which is a sphere filled with unstable circular light orbits, 
and by a secondary photon sphere on a wormhole throat 
in a black-bounce spacetime which is suggested in [F.~S.~N.~Lobo, M.~E.~Rodrigues, M.~V.~d.~S.~Silva, A.~Simpson, and M.~Visser, Phys. Rev. D {\bf 103}, 084052 (2021)]
in strong deflection limits.
There is an antiphoton sphere between the primary photon sphere and the secondary photon sphere.
If a light source and an observer are on the same side of the wormhole throat, 
in addition to an infinite number of images slightly outside of both the primary and secondary photon spheres, 
an infinite number of images formed by light rays reflected by the potential barrier near the antiphoton sphere, 
slightly inside the primary photon sphere, might be observed. 
\end{abstract}
\maketitle

\section{Introduction}
Recently, gravitational waves from binary black holes have been reported by the LIGO and Virgo Collaborations~\cite{Abbott:2016blz},
and the shadow of a supermassive black hole candidate at the center of the giant elliptical galaxy M87 has been reported by 
the Event Horizon Telescope Collaboration~\cite{Akiyama:2019cqa}. 
Phenomena in strong gravitational fields will be more important in general relativity and astrophysics than before. 
Black holes and the other compact objects such as wormholes have a sphere filled with unstable (stable) circular light orbits called a photon sphere
(an antiphoton sphere)~\cite{Perlick_2004_Living_Rev,Claudel:2000yi,Perlick:2021aok} because of their strong gravitational fields.  
There is much research on both the theoretical and observational aspects of (anti)photon spheres~\cite{Hod:2017xkz,Sanchez:1977si,Press:1971wr,Abramowicz_Prasanna_1990,Mach:2013gia,Barcelo:2000ta,Ames_1968}
and their generalized surfaces~\cite{Claudel:2000yi,Gibbons:2016isj}.
Stable circular light orbits may lead to instability of ultracompact objects because of the slow decay of linear waves~\cite{Keir:2014oka,Cardoso:2014sna,Cunha:2017qtt}.

Gravitational lensing is a useful phenomenon for surveying compact objects~\cite{Schneider_Ehlers_Falco_1992,Schneider_Kochanek_Wambsganss_2006}
and the gravitational lensing of light rays reflected by the photon sphere of compact objects has been investigated eagerly.
In 1931, Hagihara considered the images of light rays deflected by the photon sphere in the Schwarzschild spacetime~\cite{Hagihara_1931}, 
and Darwin pointed out that the images are faint in 1959~\cite{Darwin_1959}. 
The dim images by photon spheres in the Schwarzschild spacetime and in other spacetimes were revived by many researchers 
\cite{Atkinson_1965,Luminet_1979,Ohanian_1987,Nemiroff_1993,Frittelli_Kling_Newman_2000,Virbhadra_Ellis_2000,Bozza_Capozziello_Iovane_Scarpetta_2001,Bozza:2002zj,
Perlick:2003vg,Nandi:2006ds,Virbhadra:2008ws,Bozza_2010,Tsukamoto:2016zdu,Shaikh:2019jfr,Shaikh:2019itn,Tsukamoto:2020uay,Tsukamoto:2020iez,Paul:2020ufc}.

In Ref.~\cite{Bozza:2002zj}, Bozza has investigated gravitationally lensed images slightly inside a photon sphere 
in a general asymptotically flat, static, and spherically symmetric spacetime 
in a strong deflection limit $b\rightarrow b_\mathrm{m}+0$, where $b$ is the impact parameter of the light rays and $b_\mathrm{m}$ is a critical impact parameter.
The deflection angle of a light in the strong deflection limit is expressed by
\begin{eqnarray}\label{eq:al1}
\alpha &=&- \bar{a} \log \left( \frac{b}{b_\mathrm{m}}-1 \right) +\bar{b} \nonumber\\
&&+O \left( \left( \frac{b}{b_\mathrm{m}}-1 \right) \log \left( \frac{b}{b_\mathrm{m}}-1 \right) \right),
\end{eqnarray}
where $\bar{a}$ and $\bar{b}$ are described by the parameters of 
the spacetime.~\footnote{
In Ref.~\cite{Bozza:2002zj}, the order of the vanishing term is $O(b-b_\mathrm{m})$. 
However, it should be read as $O \left( \left( b/b_\mathrm{m}-1 \right) \log \left( b/b_\mathrm{m}-1 \right) \right)$ 
as discussed in Refs.~\cite{Iyer:2006cn,Tsukamoto:2016qro,Tsukamoto:2016jzh}.
}
The strong deflection limit analysis and its alternatives have been suggested~\cite{Tsukamoto:2016zdu,Shaikh:2019jfr,Shaikh:2019itn,Tsukamoto:2020uay,
Tsukamoto:2020iez,Paul:2020ufc,Bozza:2002af,Eiroa:2002mk,Petters:2002fa,Eiroa:2003jf,Bozza:2004kq,Bozza:2005tg,Bozza:2006sn,Bozza:2006nm,Iyer:2006cn,
Bozza:2007gt,Tsukamoto:2016qro,Ishihara:2016sfv,Tsukamoto:2016oca,Tsukamoto:2016jzh,Tsukamoto:2017edq,Hsieh:2021scb,Aldi:2016ntn,Takizawa:2021gdp}.

Wormholes are hypothetical objects with nontrivial topology described by general relativity~\cite{Visser_1995,Morris_Thorne_1988}, 
and their observational properties have been studied as a black hole mimicker~\cite{Damour:2007ap,Muller_2004,
James:2015ima,Ohgami:2015nra,Ohgami:2016iqm,Paul:2019trt,Kasuya:2021cpk,Nambu:2019sqn,Cardoso:2016rao}. 
Recently, Simpson and Visser~\cite{Simpson:2018tsi} have suggested a black-bounce metric which describes a (regular) black hole metric and a wormhole metric.
Gravitational lensing~\cite{Nascimento:2020ime,Ovgun:2020yuv,Tsukamoto:2020bjm,Cheng:2021hoc}
and shadows~\cite{Bronnikov:2021liv,Guerrero:2021ues} in the Simpson-Visser spacetime have been investigated.
The generalization or alternatives~\cite{Huang:2019arj,Lobo:2020ffi,Yang:2021diz,Franzin:2021vnj} 
and the rotating case~\cite{Mazza:2021rgq} of the Simpson-Visser spacetime have also been suggested. 

Recently, Shaikh \textit{et al.} have investigated gravitational lensing by light rays which fall inside a photon sphere, and which are reflected 
near a potential barrier of the antiphoton sphere of a general asymptotically flat, static, and spherically symmetric ultracompact object 
without an event horizon~\cite{Shaikh:2019itn}.  
In a strong deflection limit $b\rightarrow b_\mathrm{m}-0$, 
the deflection angle of the light rays is given by
\begin{eqnarray}\label{eq:al2}
\alpha &=&- \bar{c} \log \left( \frac{b_\mathrm{m}}{b}-1 \right) +\bar{d} \nonumber\\
&&+O \left( \left( \frac{b_\mathrm{m}}{b}-1 \right) \log \left( \frac{b_\mathrm{m}}{b}-1 \right) \right),
\end{eqnarray}
where $\bar{c}$ and $\bar{d}$ are parameters, 
and the images with the deflection angle appear slightly inside the photon sphere.

Shaikh \textit{et al.} have suggested gravitational lensing by a photon sphere on a throat 
and by another photon sphere off the throat in reflection-symmetric wormhole spacetimes~\cite{Shaikh:2018oul, Shaikh:2019jfr,Godani:2021atr},
and Gan~\textit{et al.} have investigated the shadow images of a hairy black hole with two photon spheres~\cite{Gan:2021pwu,Gan:2021xdl}.
Reflection-asymmetric thin-shell wormholes with two photon spheres of differing size, 
which could form shadow images with light rings, have been suggested 
in Refs.~\cite{Wang:2020emr,Wielgus:2020uqz,Tsukamoto:2021fpp,Guerrero:2021pxt,Peng:2021osd}.

Since shadow images and gravitational lensing by multiple photon spheres 
in strong deflection limits have been developed recently,
they have not received much research in whole parameter regions in a specific spacetime. 
To distinguish the shadow images or gravitationally lensed images by multiple photon spheres of ultracompact objects 
from the shadow image or lensed images by a photon sphere of a black hole, 
the deflection angle in the strong deflection limits (\ref{eq:al1}) and (\ref{eq:al2}) 
in the wide parameter regions of spacetimes should be studied.

In this paper, we investigate gravitational lensing in the strong deflection limits 
in a black-bounce spacetime suggested by Lobo~\textit{et al.}~\cite{Lobo:2020ffi}.
The spacetime has a mass parameter $m$ and a parameter $a$ which makes a scalar curvature regular everywhere. 
The metric corresponds to a Schwarzschild metric for $a=0$ and $m\neq 0$, it has two event horizons and two Cauchy horizons for $0<a/m<4\sqrt{3}/9$, 
it has two degenerate horizons for $a/m=4\sqrt{3}/9$, it is a traversable wormhole metric for $a/m>4\sqrt{3}/9$,
and it corresponds to an Ellis-Bronnikov wormhole metric for $a\neq 0$ and $m=0$.
The gravitationally lensed images by two photon spheres can be formed for $4\sqrt{3}/9 < a/m \leq  2\sqrt{5}/5$.

This paper is organized as follows. 
In Sec.~II, we investigate the deflection angle of a light ray in the black-bounce spacetime. 
We investigate the deflection angle and observables of gravitation lensing in strong deflection limits in Secs.~III and IV, respectively.
We shortly review gravitational lensing in a weak gravitational field in Sec.~V 
and we discuss and conclude our results in Sec VI.
In appendix~A, we consider a variable $z$ and its alternatives in a strong deflection limit analysis.
We assume that a source and an observer are on the same side of a throat if a lens is a wormhole.
We also assume that the source and the observer are far away from the photon spheres.
In this paper, we use units in which the light speed and Newton's constant are unity.

\section{Deflection angle of a light ray in a black-bounce spacetime}
Lobo \textit{et al.} have suggested a black-bounce spacetime~\cite{Lobo:2020ffi}
given by, in Buchdahl coordinates~\cite{Finch:1998,Boonserm:2007zm} with a signature $(-,+,+,+)$,
\begin{equation}\label{eq:metric}
ds^2=-A(r)dt^2+\frac{dr^2}{A(r)}+\Sigma^2(r) (d\vartheta^2+\sin^2 \vartheta d\varphi^2),
\end{equation}
where $A(r)$ and $\Sigma (r)$ are given by
\begin{equation}
A(r)=1-\frac{2M(r)}{\Sigma(r)}=1-\frac{2mr^K}{\left( r^{2N}+a^{2N} \right)^\frac{K+1}{2N}}
\end{equation}
and
\begin{equation}
\Sigma(r)=\sqrt{r^2+a^2},
\end{equation}
respectively,
and where $M(r)$ is given by
\begin{equation}
M(r)=\frac{m\Sigma(r) r^K}{\left( r^{2N}+a^{2N} \right)^\frac{K+1}{2N}}
\end{equation}
and $m$ and $a$ are nonnegative constants.
Both a time-translational Killing vector $t^\mu \partial_\mu=\partial_t$ and an axial Killing vector $\varphi^\mu \partial_\mu=\partial_\varphi$ 
are present because of the stationarity and axial symmetry of the spacetime. We assume $\vartheta=\pi/2$ without loss of generality. 
By using a wave number vector $k^\mu\equiv \dot{x}$, where the overdot denotes a differentiation with respect to an affine parameter, 
we can define the conserved energy $E\equiv -g_{\mu \nu} t^\mu k^\nu=A\dot{t}$ 
and angular momentum $L \equiv g_{\mu \nu} \varphi^\mu k^\nu=\Sigma^2 \dot{\varphi}$ of a light 
which are constant along its trajectory.
From $k^\mu k_\mu=0$, we obtain the trajectory of the light ray as
\begin{equation}\label{eq:trajectory}
-A\dot{t}^2+\frac{\dot{r}^2}{A}+\Sigma^2 \dot{\varphi}^2=0.
\end{equation}
We assume that a light ray comes from a spatial infinity $r=\infty$, reaches a closest distance $r=r_0>0$,
and returns to the same spatial infinity $r=\infty$. 
From Eq.~(\ref{eq:trajectory}), we obtain 
\begin{equation}\label{eq:trajectory0}
A_0\dot{t}^2_0=\Sigma^2_0 \dot{\varphi}^2_0.
\end{equation}
Here and hereafter, functions with the subscript $0$ denote the functions at the closest distance $r=r_0$.
We define the impact parameter of the light ray 
\begin{equation}
b(r_0)\equiv \frac{L}{E}=\frac{\Sigma_0^2 \dot{\varphi}_0}{A_0\dot{t}_0}.
\end{equation}
By using Eq.~(\ref{eq:trajectory0}), the impact parameter can be expressed by
\begin{equation}
b=\pm \sqrt{\frac{\Sigma_0^2}{A_0}}.
\end{equation}
From the definitions of the conserved energy $E$, angular momentum $L$, and impact parameter $b$, 
the equation of the trajectory~(\ref{eq:trajectory}) is rewritten by
\begin{equation}
\dot{r}^2+V(r)=0,
\end{equation}
where $V$ is an effective potential defined by
\begin{equation}
V\equiv E^2 \left( \frac{Ab^2}{\Sigma^2}-1 \right).
\end{equation}
Light rays can move only in a region of $V(r)\leq 0$. 
From the equation of the trajectory~(\ref{eq:trajectory}),
the deflection angle~$\alpha$ of the light ray is given by
\begin{equation}\label{eq:deflection}
\alpha=I(r_0)-\pi,
\end{equation}
where $I(r_0)$ is defined by
\begin{equation}\label{eq:I}
I(r_0)\equiv 2 \int^\infty_{r_0} \frac{dr}{\Sigma \sqrt{  \frac{\Sigma^2}{b^2} -A }}.
\end{equation}

When $K=0$ and $N=1$, the metric~(\ref{eq:metric}) recovers a Simpson-Visser geometry~\cite{Simpson:2018tsi}.
Lobo~\textit{et al.} pointed out that a geometry with $K=0$ and $N\geq 2$ has the same properties as the Simpson-Visser geometry.
Thus, its gravitational lensing in the strong deflection limit with $K=0$ and $N\geq 2$ 
will be very similar to the lensing in the Simpson-Visser geometry investigated in Refs.~\cite{Nascimento:2020ime,Tsukamoto:2020bjm}.

\section{Deflection angle in a strong deflection limit}
\begin{figure}[htbp]
\begin{center}
\includegraphics[width=85mm]{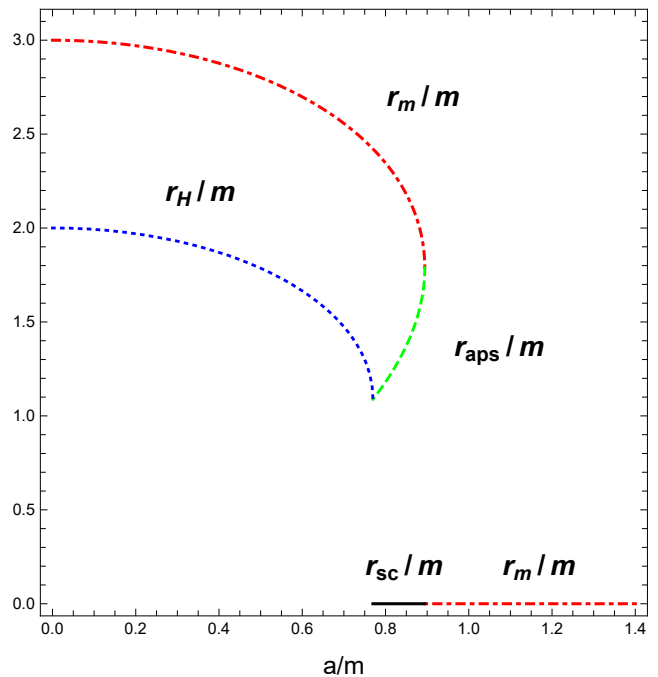}
\end{center}
\caption{The reduced radial coordinates of a primary photon sphere $r_\mathrm{m}/m$, a secondary photon sphere $r_\mathrm{sc}/m$, 
an antiphoton sphere $r_\mathrm{aps}/m$, and an event horizon $r_\mathrm{H}/m$.
A (red) dot-dashed curve at $4\sqrt{5}/5 \leq r_\mathrm{m}/m<3$ for $0\leq a/m \leq 2\sqrt{5}/5$ 
and a (red) dot-dashed line at $r_\mathrm{m}/m=0$ for $2\sqrt{5}/5 < a/m$ denote the primary photon sphere,
a (black) solid line at $r_\mathrm{sc}/m=0$ for $4\sqrt{3}/9 < a/m \leq 2\sqrt{5}/5$ denotes the secondary photon sphere,
a (green) dashed curve at $4\sqrt{6}/9 < r_{\mathrm{aps}}/m \leq 4\sqrt{5}/5$ for $4\sqrt{3}/9 < a/m < 2\sqrt{5}/5$ denotes the antiphoton sphere 
and a (blue) dotted curve at $4\sqrt{6}/9 \leq r_\mathrm{H}/m \leq 2$ for $0 \leq a/m \leq 4\sqrt{3}/9$ denotes the event horizon.}
\label{fig:radial_coorinates}
\end{figure}

In this section, we obtain the deflection angles (\ref{eq:al1}) and  (\ref{eq:al2}) in the strong deflection limits.
Hereinafter, we concentrate on a novel black-bounce spacetime with $K=2$ and $N=1$ which is suggested by Lobo \textit{et al.}~\cite{Lobo:2020ffi}.
In this case, $A(r)$ becomes
\begin{equation}
A(r)=1-\frac{2mr^2}{\left( r^{2}+a^{2} \right)^\frac{3}{2}}
\end{equation}
and an equation $A(r)=0$ has a positive solution $r=2m$ for $a=0$,
two positive and two negative solutions for $0<a/m<4\sqrt{3}/9$, 
a positive solutions $(r=4\sqrt{6}m/9)$ and a negative solution $(r=-4\sqrt{6}m/9)$ for $a/m=4\sqrt{3}/9$, 
and no real ones for $a/m>4\sqrt{3}/9$.
We concentrate on non-negative radial coordinate $r\geq 0$.
An event horizon which is the largest positive solution $r=r_\mathrm{H}$ for given $a$ is shown in Fig.~\ref{fig:radial_coorinates}.

There is a traversable wormhole throat at $r=0$ for $a/m>4\sqrt{3}/9$. 
For $a\neq 0$ and $m=0$, the metric corresponds to the Ellis-Bronnikov wormhole metric without Arnowitt-Deser-Misner~(ADM) masses~\cite{Ellis_1973,Bronnikov_1973}.
The Ellis-Bronnikov wormhole is known as the earliest passable wormhole solution of the Einstein equations with a phantom scalar field~\cite{Martinez:2020hjm}
which was obtained by Ellis and Bronnikov independently in 1973.
The metric is often included by several wormhole solutions in the vanishing ADM masses case; see Ref~\cite{Tsukamoto:2017hva} and references therein.
We note that the Ellis-Bronnikov solution is unstable~\cite{Shinkai_Hayward_2002}.
However, we might find a stable wormhole solution with the same metric 
as the Ellis-Bronnikov wormhole metric with the vanishing ADM masses, 
since the stability depends not only on the metric but also on a gravitational theory and matter source~\cite{Bronnikov:2013coa,Bronnikov:2021xao}. 

The effective potential for a light ray and its derivatives are given by
\begin{equation}
V(r)= E^2 \left\{ b^2 \left[ \frac{1}{a^2+r^2}-\frac{2m r^2}{\left(a^2+r^2\right)^\frac{5}{2}} \right] -1 \right\},
\end{equation}
\begin{equation}
V^{\prime}(r)=-\frac{2 b^2 r \left[\left(2 a^2 -3  r^2\right)m+\left(a^2+r^2\right)^\frac{3}{2}\right]}{\left(a^2+r^2\right)^\frac{7}{2}},
\end{equation}
\begin{eqnarray}
V^{\prime \prime}(r)
&=&-\frac{2 b^2}{(a^2+r^2)^\frac{9}{2}} \left[ \left(2 a^4 -21a^2r^2+12r^4 \right)m \right. \nonumber\\
&& +\left.\left(a^4-2a^2r^2-3r^4 \right)(a^2+r^2)^\frac{1}{2}\right],
\end{eqnarray}
\begin{eqnarray}
V^{\prime \prime \prime}
&=&\frac{6 b^2 r}{(a^2+r^2)^\frac{11}{2}} \left[ 5\left(4 a^4 -13a^2r^2+4r^4 \right)m \right. \nonumber\\
&& +\left. 4 (a^2-r^2) \left( a^2+r^2 \right)^\frac{3}{2}\right],
\end{eqnarray}
where the prime denotes a differentiation with respect to the Buchdahl radial coordinate~$r$.
Light rays with a critical impact parameter $b(r_0)=b(r_\mathrm{m})$ form a sphere filled with unstable circular photon orbits called photon sphere, 
which holds $V(r_\mathrm{m})=V^{\prime}(r_\mathrm{m})=0$ and $V^{\prime \prime}(r_\mathrm{m})<0$, for $a/m \neq  2\sqrt{5}/5$.
Light rays with an impact parameter $b(r_0)=b(r_{\mathrm{aps}})$ form a sphere of stable circular photon orbits called an antiphoton sphere, 
which holds $V(r_{\mathrm{aps}})=V^{\prime}(r_{\mathrm{aps}})=0$ and $V^{\prime \prime}(r_{\mathrm{aps}})>0$, for $4\sqrt{3}/9 < a/m < 2\sqrt{5}/5$.
For $a/m = 2\sqrt{5}/5$, the light ray with the critical impact parameter $b(r_0)=b(r_\mathrm{m})$ forms a marginally unstable photon sphere 
which holds $V(r_\mathrm{m})=V^{\prime}(r_\mathrm{m})=V^{\prime \prime}(r_\mathrm{m})=0$ and $V^{\prime \prime \prime}(r_\mathrm{m})<0$.
The wormhole throat at $r=0$ for $a/m>4\sqrt{3}/9$ works as a photon sphere.
Thus, for $4\sqrt{3}/9 < a/m \leq 2\sqrt{5}/5$, both the primary photon sphere at $r=r_\mathrm{m}$ and the secondary photon sphere at $r=r_\mathrm{sc}\equiv 0$ are present.~\footnote{Note that 
there is a photon sphere at $r=-r_\mathrm{m}$ for $4\sqrt{3}/9 < a/m \leq 2\sqrt{5}/5$ also.}  
We notice that the specific radius of the primary photon sphere $r_\mathrm{m}/m$ decreases monotonically 
and it changes discontinuously at $a/m = 2\sqrt{5}/5$ as $a/m$ increases.
The reduced radii of the photon spheres $r_\mathrm{m}/m$ and $r_\mathrm{sc}/m$ and the antiphoton sphere $r_{\mathrm{aps}}/m$ 
are shown in Fig.~\ref{fig:radial_coorinates}.

\subsection{Case of $a/m<4\sqrt{3}/9$}
\begin{figure}[htbp]
\begin{center}
\includegraphics[width=85mm]{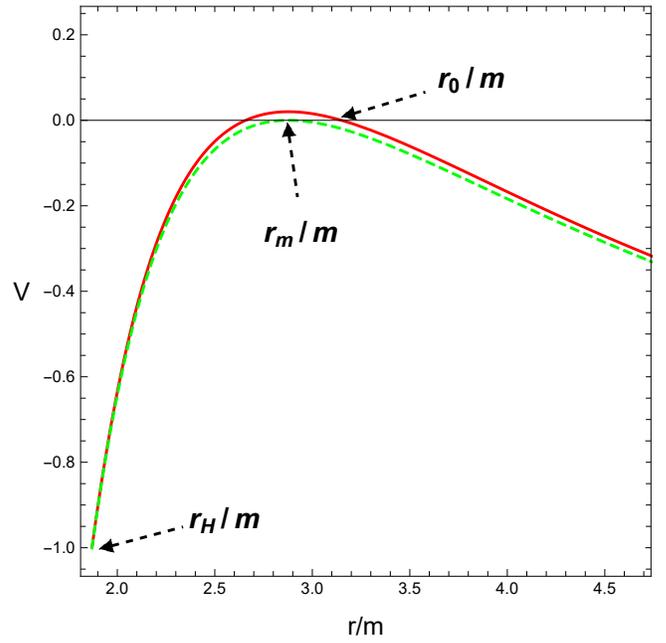}
\end{center}
\caption{Effective potentials~$V$ for images slightly outside the photon sphere with~$a/m=0.4$. Solid (red) and dashed (green) curves denote the effective potentials~$V$ 
with $b/m=1.01 b_\mathrm{m}/m=5.15$ and with $b/m=b_\mathrm{m}/m=5.10$, respectively.
The reduced radial coordinates of the photon sphere, the closest distance, and the event horizon are given by
$r_\mathrm{m}/m=2.88$, $r_\mathrm{0}/m=3.15$, and $r_\mathrm{H}/m=1.87$, respectively. $E$ is unity.}
\label{fig:V1}
\end{figure}

In the case of $a/m<4\sqrt{3}/9$, an infinite number of images are formed slightly outside of the photon sphere. 
There are no light rays coming from inside the photon sphere because of the existence of the event horizon.
The effective potential of a light ray which forms an image slightly outside the photon sphere is shown in Fig.~\ref{fig:V1}.
As shown in appendix~A, we cannot apply Bozza's formula in Ref.~\cite{Bozza:2002zj} directly for the spacetime in the Buchdahl coordinates. 
We define a variable 
\begin{equation}\label{eq:z}
z\equiv 1-\frac{r_0}{r},
\end{equation}
which is suggested in Ref.~\cite{Tsukamoto:2016jzh}.
In the case of $a/m<2\sqrt{5}/5$, 
by using the variable $z$,
we can rewrite Eq.~(\ref{eq:I}) in
\begin{equation}\label{eq:I2}
I(r_0)=\int^1_0 R(z,r_0)f(z,r_0)dz,
\end{equation}
where $R(z,r_0)$ and $f(z,r_0)$ are given by
\begin{equation}
R(z,r_0)\equiv \frac{2}{\sqrt{r_0^2+a^2(1-z)^2}},
\end{equation}
and 
\begin{eqnarray}
f(z,r_0) \equiv \frac{1}{\sqrt{h(z,r_0)}},
\end{eqnarray}
where $h(z,r_0)$ is defined by
\begin{eqnarray}
h(z,r_0) &\equiv& \frac{1}{r_0^2} \left[ \frac{r_0^2+a^2(1-z)^2}{b^2(r_0)}-(1-z)^2 \right. \nonumber\\
&&\left.+\frac{2mr_0^2(1-z)^3}{\left[ r_0^2+a^2(1-z)^2 \right]^\frac{3}{2}} \right],
\end{eqnarray}
where $b(r_0)$ can be expressed by
\begin{eqnarray}
b(r_0)=\pm\frac{\left(r_0^2+a^2\right)^\frac{5}{4}}{\left[ \left( r_0^2+a^2 \right)^\frac{3}{2}-2mr_0^2  \right]^\frac{1}{2}}.
\end{eqnarray}
Note that we obtain $h(0,r_0)=0$.
Since we are interested in $z\sim 0$, we expand $h(z,r_0)$ in the power of $z$ as
\begin{equation}
h(z,r_0)=c_1(r_0)z+c_2(r_0)z^2+O(z^3),
\end{equation}
where $c_1(r_0)$ and $c_2(r_0)$ are given by
\begin{eqnarray}
c_1(r_0)&\equiv&\frac{2 \left[ \left(2a^2-3r_0^2\right)m +\left(a^2+r_0^2\right)^\frac{3}{2} \right]}{\left(a^2+r_0^2\right)^\frac{5}{2}}, \\
c_2(r_0)&\equiv&\frac{  \left(-2a^4-11a^2r_0^2+6r_0^4 \right)m -\left(a^2+r_0^2\right)^\frac{5}{2} }{\left(a^2+r_0^2\right)^\frac{7}{2}}. \quad \quad
\end{eqnarray}
From $c_{1\mathrm{m}}\equiv c_{1}(r_\mathrm{m})=0$ and 
\begin{eqnarray}\label{eq:c2m}
c_{2\mathrm{m}}\equiv c_{2}(r_\mathrm{m})=\frac{3r^2_{\mathrm{m}} \left( r_{\mathrm{m}}^2-4a^2 \right) m }{\left( a^2+r_{\mathrm{m}}^2 \right)^\frac{7}{2}},
\end{eqnarray}
the term $I(r_0)$ diverges logarithmically in the strong deflection limit $r_0 \rightarrow r_\mathrm{m}+0$. 
Here and hereinafter, functions with the subscript $\mathrm{m}$ denote the functions at the photon sphere $r_0=r_\mathrm{m}$.
We expand $c_1(r_0)$ and $b$ in the power of $r_0-r_{\mathrm{m}}>0$ as
\begin{eqnarray}\label{eq:c1ex}
c_1(r_0)=c'_{1\mathrm{m}}(r_0-r_{\mathrm{m}})+O\left((r_0-r_{\mathrm{m}})^2\right)
\end{eqnarray}
and
\begin{equation}
b(r_0)=b_{\mathrm{m}}+\frac{1}{2}b^{\prime \prime}_{\mathrm{m}} \left( r_0-r_{\mathrm{m}} \right)^2+O\left( \left( r_0-r_{\mathrm{m}} \right)^3 \right),
\end{equation}
respectively, where 
$c^\prime_{1\mathrm{m}}$, $b_{\mathrm{m}}$, and $b^{\prime \prime}_{\mathrm{m}}$ are given by
\begin{eqnarray}
c^\prime_{1\mathrm{m}}&=&\frac{6r_{\mathrm{m}} \left( r_{\mathrm{m}}^2-4a^2 \right) m }{\left( a^2+r_{\mathrm{m}}^2 \right)^\frac{7}{2}},
\end{eqnarray}
\begin{eqnarray}
b_{\mathrm{m}}=\pm \frac{\left(r_{\mathrm{m}}^2+a^2\right)^\frac{5}{4}}{\left( r_{\mathrm{m}}^2-2a^2 \right)^\frac{1}{2} m^\frac{1}{2}},
\end{eqnarray}
and
\begin{eqnarray}\label{eq:bppm}
b^{\prime \prime}_{\mathrm{m}}=\pm\frac{r_{\mathrm{m}}\left(r_{\mathrm{m}}^2+a^2\right)^\frac{11}{4}c_{1\mathrm{m}}^{\prime}}{2\left( r_{\mathrm{m}}^2-2a^2 \right)^\frac{3}{2} m^\frac{3}{2}},
\end{eqnarray}
respectively.
Notice that $b^\prime_{\mathrm{m}}=0$ because $b^\prime$ is obtained as
\begin{eqnarray}
b^\prime&=&\pm\frac{r_0\left(r_0^2+a^2\right)^\frac{11}{4}c_1}{2\left[ \left( r_0^2+a^2 \right)^\frac{3}{2}-2mr_0^2  \right]^\frac{3}{2}}.
\end{eqnarray}
From now on, we concentrate on the positive impact parameter $b$ unless we explicitly state that we are paying attention to the negative one.

We separate $I(r_0)$ into a divergent part~$I_\mathrm{D}$ and a regular part~$I_\mathrm{R}$.
In the case of $a/m<2\sqrt{5}/5$, we define $I_\mathrm{D}$ as 
\begin{equation}
I_\mathrm{D}\equiv \int^1_0 R(0,r_{\mathrm{m}}) f_\mathrm{D}(z,r_0)dz,
\end{equation}
where $f_\mathrm{D}(z,r_0)$ is defined as 
\begin{equation}
f_\mathrm{D}(z,r_0) \equiv \frac{1}{\sqrt{c_1(r_0)z+c_2(r_0)z^2}}
\end{equation}
and $R(0,r_{\mathrm{m}})$ is obtained as 
\begin{equation}\label{eq:R0rm}
R(0,r_{\mathrm{m}})=\frac{2}{\sqrt{r_{\mathrm{m}}^2+a^2}}.
\end{equation}
As shown by Bozza~\cite{Bozza:2002zj}, 
we can integrate $I_\mathrm{D}$ as
\begin{equation}
I_\mathrm{D}= \frac{2R(0,r_{\mathrm{m}})}{\sqrt{c_2(r_0)}} \log \frac{\sqrt{c_2(r_0)}+\sqrt{c_1(r_0)+c_2(r_0)}}{\sqrt{c_1(r_0)}}.
\end{equation}
By using Eqs.~(\ref{eq:c2m})-(\ref{eq:bppm}) and (\ref{eq:R0rm}), 
$I_\mathrm{D}$ in the strong deflection limit $r_0 \rightarrow r_{\mathrm{m}}+0$ or $b \rightarrow b_{\mathrm{m}}+0$ is expressed by
\begin{equation}
I_\mathrm{D}
= -\bar{a} \log \left( \frac{b}{b_{\mathrm{m}}}-1 \right) 
+\bar{a} \log \frac{6r_{\mathrm{m}}^4(r_{\mathrm{m}}^2-4a^2)}{(a^2+r_{\mathrm{m}}^2)^2(r_{\mathrm{m}}^2-2a^2)},
\end{equation}
where $\bar{a}$ is given by
\begin{equation}
\bar{a}= \frac{\left( a^2+r_{\mathrm{m}}^2 \right)^\frac{5}{4}}{r_{\mathrm{m}} \sqrt{3(r_{\mathrm{m}}^2-4a^2)m}}.
\end{equation}

\begin{figure}[htbp]
\begin{center}
\includegraphics[width=85mm]{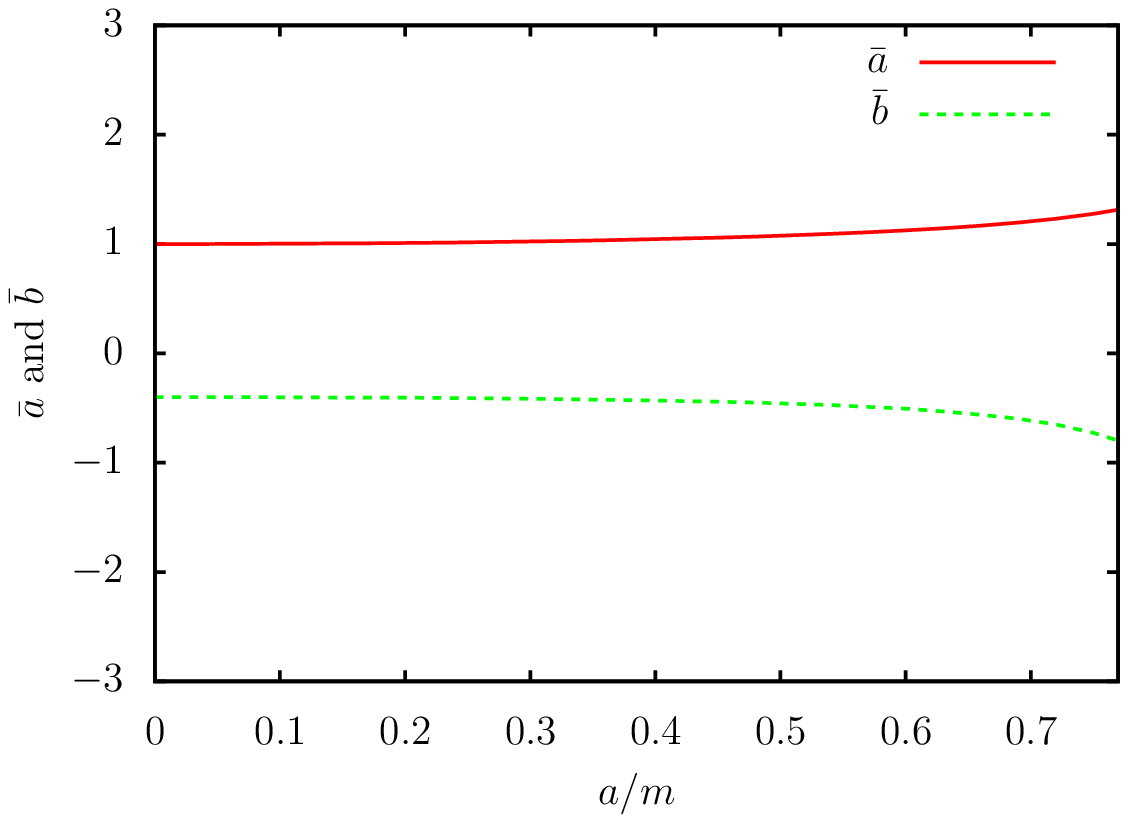}
\end{center}
\caption{Parameters $\bar{a}$ and $\bar{b}$ of the deflection angle~(\ref{eq:al1}) in the strong deflection limit  
for the images slightly outside of the photon sphere in the case of $a/m<4\sqrt{3}/9$.
Solid (red) and dashed (green) curves denote $\bar{a}$ and $\bar{b}$, respectively.}
\label{fig:abcd1}
\end{figure}

The regular part $I_\mathrm{R}$ is defined by
\begin{equation}
I_\mathrm{R}(r_0)\equiv \int^1_0 g(z,r_0) dz, 
\end{equation}
where $g(z,r_0)$ is given by
\begin{equation}
g(z,r_0)\equiv R(z,r_0)f(z,r_0)-R(0,r_{\mathrm{m}})f_\mathrm{D}(z,r_0)
\end{equation}
and it can be expanded in the power of $r_0-r_{\mathrm{m}}$ as 
\begin{equation}
I_\mathrm{R}(r_0)=\sum^\infty_{j=0} \frac{1}{j!}(r_0-r_{\mathrm{m}})^j \int^1_0 \left. \frac{\partial^j g}{\partial r_0^j} \right|_{r_0=r_{\mathrm{m}}} dz
\end{equation}
and we are only interested in the first term
\begin{equation}
I_\mathrm{R}=\int^1_0 g(z,r_{\mathrm{m}}) dz.
\end{equation}
Usually, $I_\mathrm{R}$ is calculated numerically, and $\bar{b}$ is obtained as
\begin{equation}
\bar{b}= \bar{a} \log \frac{6r_{\mathrm{m}}^4(r_{\mathrm{m}}^2-4a^2)}{(a^2+r_{\mathrm{m}}^2)^2(r_{\mathrm{m}}^2-2a^2)}+I_\mathrm{R}-\pi.
\end{equation}
The parameters $\bar{a}$ and $\bar{b}$ of the deflection angle (\ref{eq:al1}) in the strong deflection limit 
are plotted in Fig.~\ref{fig:abcd1}.

In the Schwarzschild spacetime case, i.e., $a=0$, we get $b_{\mathrm{m}}=3\sqrt{3}m$, $r_{\mathrm{m}}=3m$, 
\begin{equation}
I_\mathrm{D}= -\log \left( \frac{b}{b_{\mathrm{m}}}-1 \right)+ \log 6,
\end{equation}
and
\begin{equation}
I_\mathrm{R}=\int^1_0  \left( \frac{2}{z\sqrt{1-\frac{2z}{3}}}- \frac{2}{z} \right) dz =2\log[6(2-\sqrt{3})].
\end{equation}
Thus, we obtain $\bar{a}=1$ and 
\begin{equation}
\bar{b}=\log [216(7-4\sqrt{3})] -\pi \sim -0.40
\end{equation}
which recovers the known results of the Schwarzschild spacetime in Refs.~\cite{Darwin_1959,Bozza_Capozziello_Iovane_Scarpetta_2001,Bozza:2002zj,Iyer:2006cn}.

\subsection{Case of $4\sqrt{3}/9<a/m<2\sqrt{5}/5$}
For $4\sqrt{3}/9<a/m<2\sqrt{5}/5$, a wormhole has the primary photon sphere off a throat and the secondary photon sphere on the throat,
and an antiphoton sphere between them. 
Three sets of infinite numbers of images are formed as shown below.

\subsubsection{Images slightly outside of the primary photon sphere~(ISOP)}
\begin{figure}[htbp]
\begin{center}
\includegraphics[width=85mm]{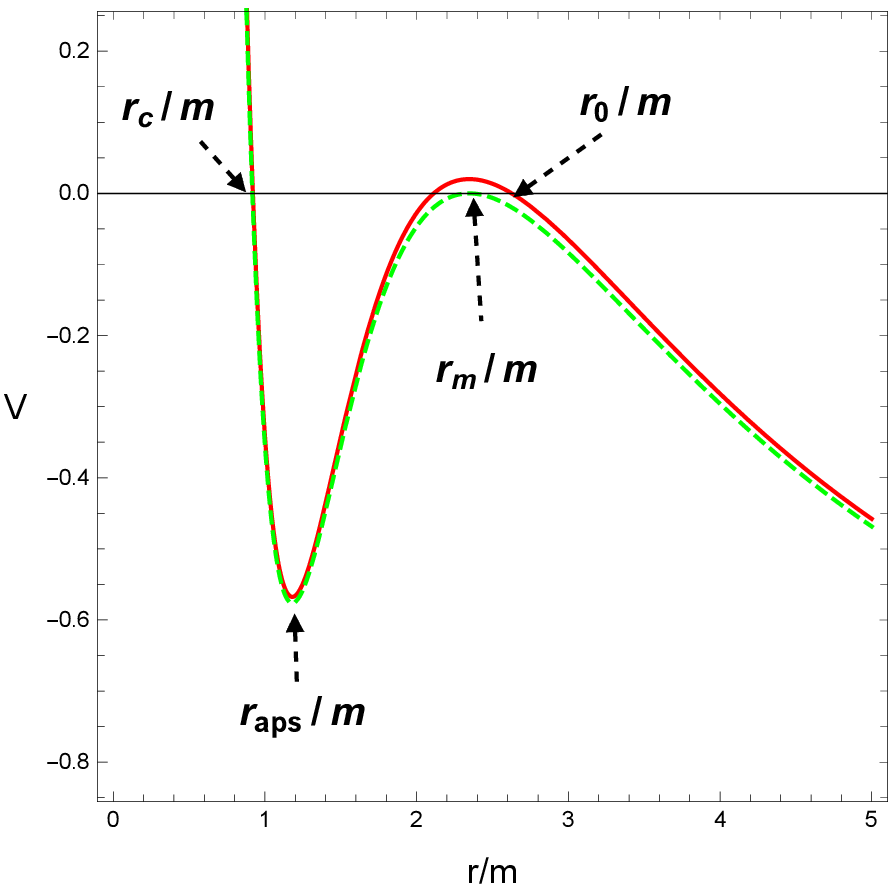}
\end{center}
\caption{Effective potentials~$V$ for images slightly outside of the primary photon sphere with~$a/m=0.8$. Solid (red) and dashed (green) curves denote the effective potentials~$V$ 
with $b/m=1.01 b_\mathrm{m}/m=4.71$ and with $b/m=b_\mathrm{m}/m=4.76$, respectively.
The reduced radial coordinates of the primary photon sphere, the antiphoton sphere, the closest distance of a light with $b=1.01b_\mathrm{m}$, 
and the smaller positive root of the effective potential with $b=b_\mathrm{m}$ are given by
$r_\mathrm{m}/m=2.35$, $r_\mathrm{aps}/m=1.18$, $r_\mathrm{0}/m=2.63$, and $r_\mathrm{c}/m=0.917$, respectively. $E$ is unity.}
\label{fig:V2}
\end{figure}

\begin{figure}[htbp]
\begin{center}
\includegraphics[width=85mm]{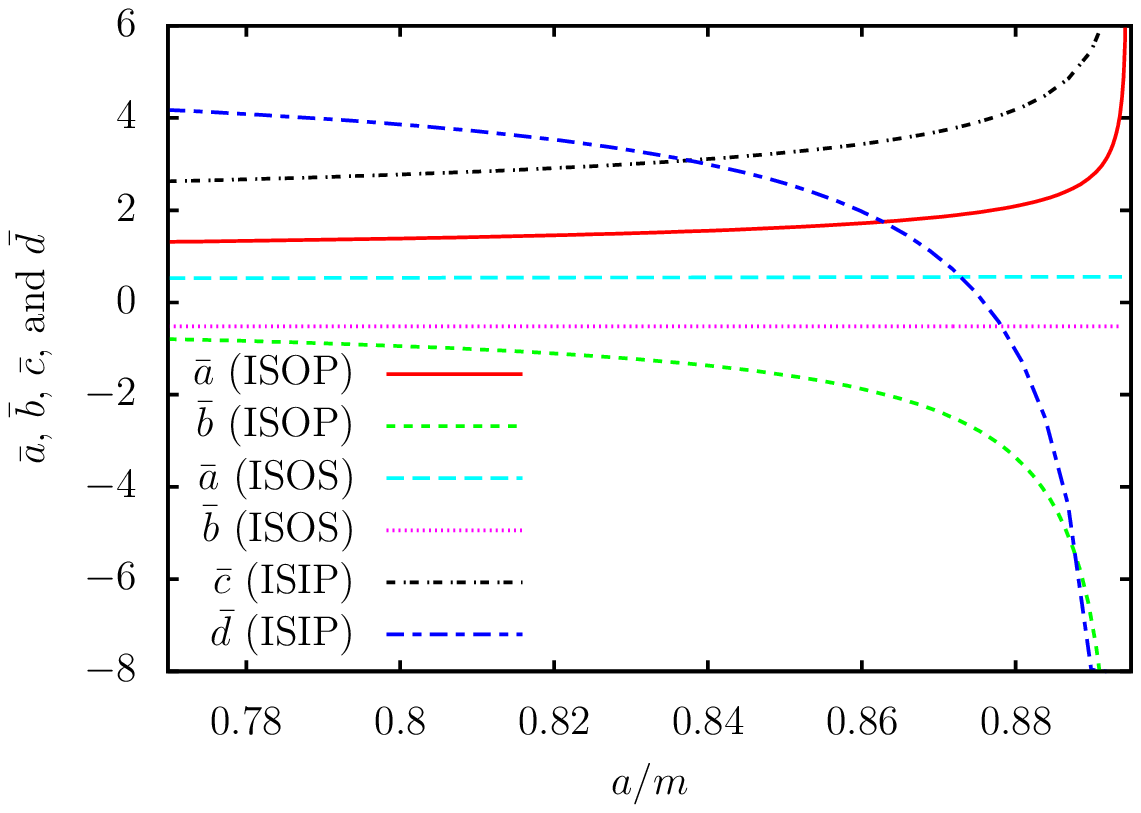}
\end{center}
\caption{Parameters $\bar{a}$, $\bar{b}$, $\bar{c}$, and $\bar{d}$ in the deflection angles~(\ref{eq:al1}) and (\ref{eq:al2}) in the strong deflection limits for $4\sqrt{3}/9<a/m<2\sqrt{5}/5$.  
Solid red and dashed green curves denote $\bar{a}$ and $\bar{b}$ for the images slightly outside of the primary photon sphere~(ISOP), respectively.
Long-dashed cyan and dotted magenta curves denote $\bar{a}$ and $\bar{b}$ for the images slightly outside of the secondary photon sphere~(ISOS), respectively.
Dot-dashed black and long-dashed-short-dashed blue curves denote $\bar{c}$ and $\bar{d}$ for the images slightly inside the primary photon sphere~(ISIP), respectively.
}
\label{fig:abcd2}
\end{figure}

Figure~\ref{fig:V2} shows the effective potential of a light ray which makes an image slightly outside of the primary photon sphere.
We can use the formulas in Sec.~III-A.
Parameters $\bar{a}$ and $\bar{b}$ in the deflection angle~(\ref{eq:al1}) for the images slightly outside of the primary photon sphere 
in the strong deflection limit are shown in Fig.~\ref{fig:abcd2}.

\subsubsection{Images slightly inside the primary photon sphere~(ISIP)}
\begin{figure}[htbp]
\begin{center}
\includegraphics[width=85mm]{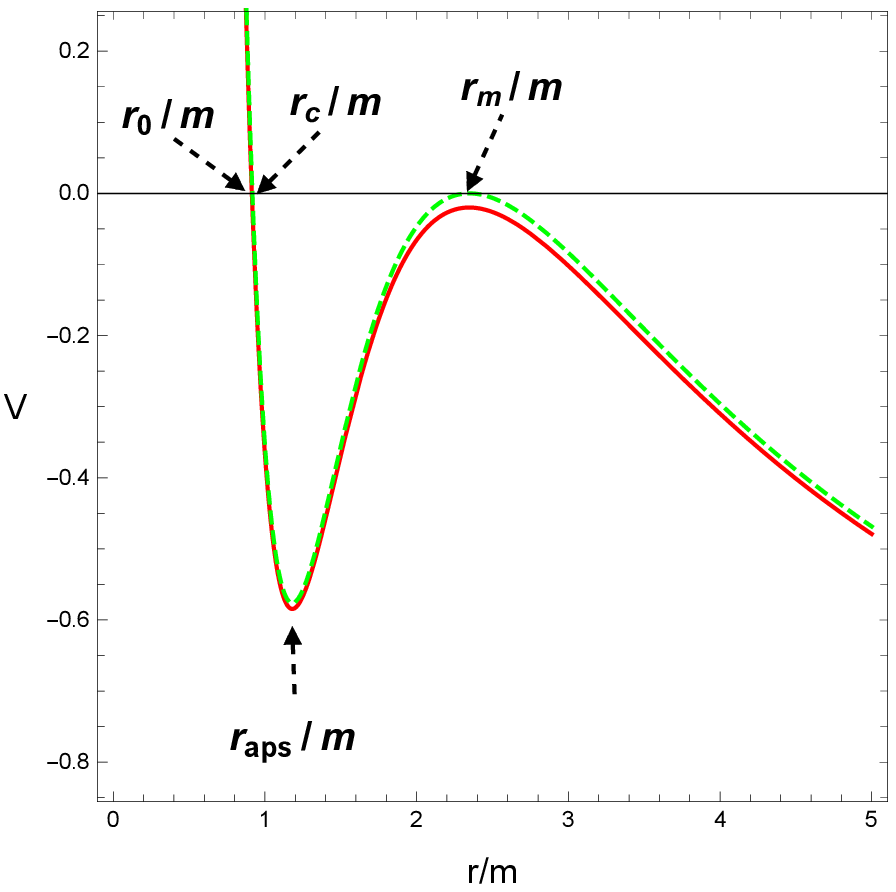}
\end{center}
\caption{
Effective potentials~$V$ for images slightly inside the primary photon sphere with~$a/m=0.8$. Solid (red) and dashed (green) curves denote the effective potentials~$V$ 
with $b/m=0.99 b_\mathrm{m}/m=4.66$ and with $b/m=b_\mathrm{m}/m=4.76$, respectively.
The reduced radial coordinates of the primary photon sphere, the antiphoton sphere,  the closest distance of the light with $b=0.99b_\mathrm{m}$, 
and the smaller positive root of the effective potential with $b=b_\mathrm{m}$ are given by
$r_\mathrm{m}/m=2.35$, $r_\mathrm{aps}/m=1.18$, $r_\mathrm{0}/m=0.913$, and $r_\mathrm{c}/m=0.917$, respectively. $E$ is unity.
}
\label{fig:V3}
\end{figure}

Light rays which are reflected near the antiphoton sphere form images slightly inside the primary photon sphere.
The effective potential of such a light ray is shown in Fig.~\ref{fig:V3}.
We introduce a variable $\mathrm{z}$, which is suggested in Ref.~\cite{Shaikh:2019itn}:  
\begin{equation}
\mathrm{z}\equiv 1-\frac{r_\mathrm{m}}{r}.
\end{equation}
The smaller positive root of the effective potential caused by the antiphoton sphere with $b=b_\mathrm{m}$ is denoted by $r=r_\mathrm{c}$.
Note that $b_\mathrm{c}\equiv b(r_\mathrm{c})=b_\mathrm{m}.$
Here and hereinafter, functions with the subscript $\mathrm{c}$ denote the function at $r_0=r_\mathrm{c}$.
By using $\mathrm{z}$, we rewrite Eq.~(\ref{eq:I}) in
\begin{equation}
I(r_0)=\int^1_{\beta(r_0)} F(\mathrm{z},r_0)d\mathrm{z},
\end{equation}
where $\beta(r_0)$ and $F(\mathrm{z},r_0)$ are given by
\begin{equation}
\beta\equiv 1-\frac{r_\mathrm{m}}{r_0},
\end{equation}
and 
\begin{eqnarray}
F(\mathrm{z},r_0) = \frac{2}{\sqrt{H(\mathrm{z},r_0)}},
\end{eqnarray}
where $H(\mathrm{z},r_0)$ is defined by
\begin{eqnarray}
H(\mathrm{z},r_0) &=&  \frac{r_\mathrm{m}^2+a^2(1-\mathrm{z})^2}{r_\mathrm{m}^2 } \left[ \frac{r_\mathrm{m}^2+a^2(1-\mathrm{z})^2}{b^2(r_0)} \right. \nonumber\\
&&\left.-(1-\mathrm{z})^2+\frac{2mr_\mathrm{m}^2(1-\mathrm{z})^3}{\left[ r_\mathrm{m}^2+a^2(1-\mathrm{z})^2 \right]^\frac{3}{2}} \right].
\end{eqnarray}
We expand $H(\mathrm{z},r_0)$ around $\mathrm{z}=0$ as
\begin{equation}
H(\mathrm{z},r_0)=c_3(r_0)+c_4(r_0)\mathrm{z}+c_5(r_0)\mathrm{z}^2+O(\mathrm{z}^3),
\end{equation}
where $c_3(r_0)$, $c_4(r_0)$, and $c_5(r_0)$ are given by
\begin{eqnarray}
c_3(r_0)&\equiv& \frac{\left( r_\mathrm{m}^2+a^2 \right)^2}{r_\mathrm{m}^2 b_\mathrm{m}^2} \left( \frac{b_\mathrm{m}^2}{b^2}-1 \right),\\
c_4(r_0)&\equiv& -\frac{4a^2 \left( r_\mathrm{m}^2+a^2 \right)}{r_\mathrm{m}^2 b_\mathrm{m}^2} \left( \frac{b_\mathrm{m}^2}{b^2}-1 \right),\\
c_5(r_0)&\equiv& \frac{2a^2 \left( r_\mathrm{m}^2+3a^2 \right)}{r_\mathrm{m}^2 b_\mathrm{m}^2} \left( \frac{b_\mathrm{m}^2}{b^2}-1 \right) \nonumber\\
&&+\frac{3 m r_\mathrm{m}^2 \left( r_\mathrm{m}^2-4a^2 \right)}{\left( r_\mathrm{m}^2+a^2 \right)^\frac{5}{2}}. 
\end{eqnarray}
Note that we obtain, in a strong deflection limit $r_0 \rightarrow r_\mathrm{c}-0$ or $b \rightarrow b_\mathrm{c}-0=b_\mathrm{m}-0$, 
\begin{eqnarray}
c_3(r_0)&\rightarrow& +0, \\
c_4(r_0)&\rightarrow& -0, \\
\beta(r_0)&\rightarrow&1-\frac{r_\mathrm{m}}{r_\mathrm{c}}<0, \\ \label{eq:c11}
c_5(r_0)&\rightarrow& \frac{3 m r_\mathrm{m}^2 \left( r_\mathrm{m}^2-4a^2 \right)}{\left( r_\mathrm{m}^2+a^2 \right)^\frac{5}{2}}>0.
\end{eqnarray}
Thus, the term $I(r_0)$ diverges logarithmically in the strong deflection limit $r_0 \rightarrow r_\mathrm{c}-0$.  
We expand $c_3(r_0)$ and $b(r_0)$ in the power of $r_0-r_{\mathrm{c}}<0$ as
\begin{eqnarray}
c_3(r_0)=c'_{3\mathrm{c}}(r_0-r_{\mathrm{c}})+O\left((r_0-r_{\mathrm{c}})^2\right)
\end{eqnarray}
and
\begin{equation}
b(r_0)=b_{\mathrm{c}}+b^{\prime}_{\mathrm{c}} \left( r_0-r_{\mathrm{c}} \right)+O\left( \left( r_0-r_{\mathrm{c}} \right)^2 \right),
\end{equation}
respectively, where $c^\prime_{3\mathrm{c}}$, $b_{\mathrm{c}}$, and $b^{\prime \prime}_{\mathrm{c}}$ are given by
\begin{eqnarray}
c^\prime_{3\mathrm{c}}&=&-\frac{2\left( r_\mathrm{m}^2+a^2 \right)^2 b^{\prime}_\mathrm{c}}{r_{\mathrm{m}}^2 b_{\mathrm{m}}^3}, \\\label{eq:bc}
b_{\mathrm{c}}&=&b_\mathrm{m}=\frac{\left(r_{\mathrm{m}}^2+a^2\right)^\frac{5}{4}}{\left( r_{\mathrm{m}}^2-2a^2 \right)^\frac{1}{2} m^\frac{1}{2}},
\end{eqnarray}
and
\begin{equation}
b_\mathrm{c}^\prime=\frac{r_\mathrm{c}\left(r_\mathrm{c}^2+a^2\right)^\frac{1}{4} \left[ (-3r_\mathrm{c}^2+2a^2)m+(r_\mathrm{c}^2+a^2)^\frac{3}{2} \right]}{\left[ \left( r_\mathrm{c}^2+a^2 \right)^\frac{3}{2}-2mr_\mathrm{c}^2  \right]^\frac{3}{2}},
\end{equation}
respectively.
Therefore, we obtain 
\begin{eqnarray}\label{eq:c3c}
c_{3\mathrm{c}}= \frac{2\left( r_\mathrm{m}^2+a^2 \right)^2}{r_\mathrm{m}^2 b_\mathrm{m}^2} \left( 1- \frac{b}{b_\mathrm{m}} \right).
\end{eqnarray}

We define the divergent term $I_\mathrm{D}$ by
\begin{equation}
I_\mathrm{D}\equiv \int^1_{\beta(r_0)} F_\mathrm{D}(\mathrm{z},r_0)d\mathrm{z},
\end{equation}
where $F_\mathrm{D}(\mathrm{z},r_0)$ is defined as 
\begin{equation}
F_\mathrm{D}(\mathrm{z},r_0) \equiv \frac{2}{\sqrt{c_3(r_0)+c_4(r_0)\mathrm{z}+c_5(r_0)\mathrm{z}^2}}.
\end{equation}
We can integrate $I_\mathrm{D}$ as~\cite{Shaikh:2019itn}
\begin{equation}
I_\mathrm{D}= \frac{2}{\sqrt{c_5}} \log \frac{c_4+2c_5+2\sqrt{c_5(c_3+c_4+c_5)}}{c_4+2c_5\beta+2\sqrt{c_5\left( c_3+c_4\beta+c_5\beta^2\right)}}.
\end{equation}
In the strong deflection limit $r_0 \rightarrow r_\mathrm{c}-0$ or $b \rightarrow b_\mathrm{c}-0=b_{\mathrm{m}}-0$,
by using the approximations
\begin{equation}
\sqrt{c_{5\mathrm{c}}\left( c_{3\mathrm{c}}+c_{4\mathrm{c}}\beta_{\mathrm{c}}+c_{5\mathrm{c}}\beta^2_{\mathrm{c}}\right)}
\sim -c_{5\mathrm{c}} \beta_{\mathrm{c}} \left( 1+ \frac{c_{3\mathrm{c}}+c_{4\mathrm{c}}\beta_{\mathrm{c}}}{2 c_{5\mathrm{c}} \beta^2_{\mathrm{c}}} \right)
\end{equation}
and 
\begin{equation}
\frac{b_\mathrm{m}^2}{b^2}-1 \sim 2 \left( 1-\frac{b}{b_\mathrm{m}} \right)
\end{equation}
and by using $c_{3\mathrm{c}}$~(\ref{eq:c3c}),
we obtain
\begin{eqnarray}
I_\mathrm{D}
&=&\bar{c} \log \left( -\frac{4c_{5\mathrm{c}} \beta_\mathrm{c}}{c_{3\mathrm{c}}} \right) \nonumber\\
&=&-\bar{c} \log \left( 1-\frac{b}{b_\mathrm{m}} \right) \nonumber\\
&&+\bar{c} \log \left(  \frac{6r_\mathrm{m}^4(r_\mathrm{m}^2-4a^2)}{(r_\mathrm{m}^2+a^2)^2 (r_\mathrm{m}^2-2a^2)} \left( \frac{r_\mathrm{m}}{r_\mathrm{c}}-1 \right) \right), \nonumber\\
\end{eqnarray}
where $\bar{c}$ is given by
\begin{equation}
\bar{c}\equiv \frac{2\left( a^2+r_{\mathrm{m}}^2 \right)^\frac{5}{4}}{r_{\mathrm{m}} \sqrt{3(r_{\mathrm{m}}^2-4a^2)m}}= 2\bar{a}.
\end{equation}

The regular part $I_\mathrm{R}$ is defined by
\begin{equation}
I_\mathrm{R}(r_0)\equiv \int^1_{\beta(r_0)} G(\mathrm{z},r_0) d\mathrm{z}, 
\end{equation}
where $G(\mathrm{z},r_0)$ is given by
\begin{equation}
G(\mathrm{z},r_0)\equiv F(\mathrm{z},r_0)-F_\mathrm{D}(\mathrm{z},r_0).
\end{equation}
We expand $I_\mathrm{R}(r_0)$ in the power of $r_0-r_{\mathrm{c}}$ as 
\begin{equation}
I_\mathrm{R}(r_0)=\sum^\infty_{j=0} \frac{1}{j!}(r_0-r_{\mathrm{c}})^j \int^1_{\beta(r_\mathrm{c})} \left. \frac{\partial^j G}{\partial r_0^j} \right|_{r_0=r_{\mathrm{c}}} d\mathrm{z}
\end{equation}
and the first term 
\begin{equation}
I_\mathrm{R}=\int^1_{\beta(r_\mathrm{c})} G(\mathrm{z},r_{\mathrm{c}}) d\mathrm{z}
\end{equation}
is calculated numerically.
The term $\bar{d}$ is expressed as
\begin{equation}
\bar{d}=\bar{c} \log \left(  \frac{6r_\mathrm{m}^4(r_\mathrm{m}^2-4a^2)}{(r_\mathrm{m}^2+a^2)^2 (r_\mathrm{m}^2-2a^2)} \left( \frac{r_\mathrm{m}}{r_\mathrm{c}}-1 \right) \right) +I_\mathrm{R}-\pi. 
\end{equation}
Parameters $\bar{c}$ and $\bar{d}$ in the deflection angle~(\ref{eq:al2}) for the images slightly inside the primary photon sphere 
in the strong deflection limit~$b \rightarrow b_{\mathrm{m}}-0$ are shown in Fig.~\ref{fig:abcd2}.

\subsubsection{Images slightly outside of the secondary photon sphere~(ISOS)}
\begin{figure}[htbp]
\begin{center}
\includegraphics[width=85mm]{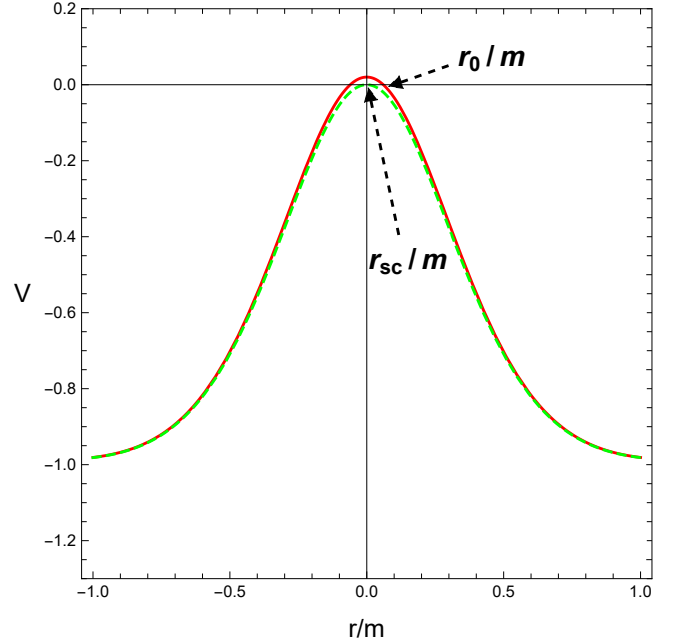}
\end{center}
\caption{
Effective potentials~$V$ for images slightly outside of the secondary photon sphere with~$a/m=0.8$. Solid (red) and dashed (green) curves denote the effective potentials~$V$ 
with $b/m=1.01 b_\mathrm{sc}/m=0.808$ and with $b/m=b_\mathrm{sc}/m=0.8$, respectively.
The reduced radial coordinates of the secondary photon sphere and the closest distance of the light with $b=1.01 b_\mathrm{sc}$ are given by
$r_\mathrm{sc}/m=0$ and $r_\mathrm{0}/m=0.0604$, respectively. $E$ is unity.}
\label{fig:V4}
\end{figure}

For $4\sqrt{3}/9<a/m\leq 2\sqrt{5}/5$, 
the wormhole throat works as the secondary photon sphere at $r=r_{\mathrm{sc}}=0$ and it reflects light rays to form images slightly outside of the throat.
The effective potential of the light ray which is reflected by the secondary photon sphere is shown in Fig.~\ref{fig:V4}.

The variable $z$~(\ref{eq:z}) in a strong deflection limit $r_0\rightarrow r_{\mathrm{sc}}+0=+0$
would not work well, since the variable $z$ does not depend on $r$ in the limit. 
Thus, we use not the Buchdahl radial coordinate $r$ but a standard radial coordinate $\rho\equiv \sqrt{r^2+a^2}$. 
Under the standard radial coordinate, 
the line element is expressed as 
\begin{eqnarray}\label{eq:metric2}
ds^2&=&-\left[1-\frac{2m(\rho^2-a^2)}{\rho^3}\right]dt^2 \nonumber\\
&&+\frac{d\rho^2}{\left[1-\frac{2m(\rho^2-a^2)}{\rho^3}\right] \left( 1-\frac{a^2}{\rho^2}\right)} 
+\rho^2 (d\vartheta^2+\sin^2 \vartheta d\varphi^2), \nonumber\\
\end{eqnarray}
Notice that the radius of the secondary photon sphere is given by $\rho=\rho_{\mathrm{sc}}=a$.
Here and hereinafter, functions with the subscript $\mathrm{sc}$ denote the function at $r_0=r_\mathrm{sc}$.
By using the alternative variable $\bar{z}$ defined by
\begin{equation}\label{eq:barz}
\bar{z}\equiv 1-\frac{\rho_0}{\rho},
\end{equation}
the term $I$~(\ref{eq:I}) can be rewritten by
\begin{equation}
I(\rho_0)=\int^1_0 k(\bar{z},\rho_0)d\bar{z},
\end{equation}
where $k(\bar{z},\rho_0)$ is given by
\begin{eqnarray}
k(\bar{z},\rho_0)\equiv \frac{2}{\sqrt{J(\bar{z},\rho_0)}}, 
\end{eqnarray}
where $J(\bar{z},\rho_0)$ is defined by
\begin{eqnarray}
J(\bar{z},\rho_0)&\equiv& \left[ 1-\frac{a^2(1-\bar{z})^2}{\rho^2_0} \right] \left\{ (2-\bar{z})\bar{z} \right. \nonumber\\
&& +\left.  \frac{2m}{\rho^3_0}\left[ a^2-\rho^2_0+\rho^2_0(1-\bar{z})^3-a^2(1-\bar{z})^5 \right] \right\} \nonumber\\
\end{eqnarray}
and we expand $J(\bar{z},\rho_0)$ around $\bar{z}=0$ as 
\begin{eqnarray}
J(\bar{z},\rho_0)=c_6(\rho_0)\bar{z}+c_7(\rho_0)\bar{z}^2+O(\bar{z}^3),
\end{eqnarray}
where $c_6(\rho_0)$ and $c_7(\rho_0)$ are given by
\begin{eqnarray}
c_6(\rho_0)&\equiv&\frac{2\left( \rho_0^2-a^2 \right)}{\rho^2_0} \left[ 1+\frac{m\left(5a^2-3\rho_0^2\right)}{\rho^3_0} \right], \\
c_7(\rho_0)&\equiv&\frac{-\rho_0^2+5a^2}{\rho_0^2} +\frac{m\left( 6\rho_0^4-38\rho_0^2a^2+40a^4 \right)}{\rho_0^5}.\quad 
\end{eqnarray}
From 
\begin{eqnarray}
c_{6\mathrm{sc}}&=&0, \\
c_{7\mathrm{sc}}&=&\frac{4(a+2m)}{a},
\end{eqnarray}
$I(\rho)$ diverges logarithmically in the strong deflection limit $\rho_0 \rightarrow \rho_{\mathrm{sc}}+0$.
We separate $I(\rho)$ as a divergent part $I_\mathcal{D}$ and a regular part $I_\mathcal{R}$.
We define $I_\mathcal{D}$ as
\begin{eqnarray}
I_\mathcal{D}
\equiv \int^1_0 k_D(\bar{z},\rho_0)d\bar{z},
\end{eqnarray}
where $k_\mathcal{D}(\bar{z},\rho_0)$ is 
\begin{equation}
k_\mathcal{D}(\bar{z},\rho_0)\equiv \frac{2}{\sqrt{c_6(\rho_0)\bar{z}+c_7(\rho_0)\bar{z}^2}}
\end{equation}
and we obtain $I_\mathcal{D}$ as 
\begin{eqnarray}
I_\mathcal{D}
= \frac{4}{\sqrt{c_7}} \log \frac{\sqrt{c_7}+\sqrt{c_6+c_7}}{\sqrt{c_6}}.
\end{eqnarray}
By using 
\begin{equation}
c_6(\rho_0)=\left.\frac{dc_6}{d\rho_0}\right|_{\rho_0=\rho_{\mathrm{sc}}}(\rho_0-\rho_{\mathrm{sc}})+O\left((\rho_0-\rho_{\mathrm{sc}})^2\right), 
\end{equation}
where 
\begin{equation}
\left.\frac{dc_{6}}{d\rho_0}\right|_{\rho_0=\rho_{\mathrm{sc}}}=\frac{4(a+2m)}{a^2},
\end{equation}
and 
\begin{equation}
b=b_{\mathrm{sc}}+\left.\frac{db}{d\rho_0}\right|_{\rho_0=\rho_{\mathrm{sc}}}(\rho_0-\rho_{\mathrm{sc}})+O\left((\rho_0-\rho_{\mathrm{sc}})^2\right), 
\end{equation}
where $b_{\mathrm{sc}}=a$ and 
\begin{eqnarray}
\left.\frac{db}{d\rho_0}\right|_{\rho_0=\rho_{\mathrm{sc}}}=\frac{a+2m}{a},
\end{eqnarray}
we obtain the divergent part $I_\mathcal{D}$ as
\begin{eqnarray}
I_\mathcal{D}=-\bar{a} \log \left( \frac{b}{b_{\mathrm{sc}}}-1 \right)+\bar{a} \log \frac{4(a+2m)}{a},
\end{eqnarray}
where
\begin{equation}
\bar{a}=\sqrt{\frac{a}{a+2m}}.
\end{equation}

The regular part $I_\mathcal{R}$ is defined by
\begin{eqnarray}
I_\mathcal{R}\equiv  \int^1_0 k_\mathcal{R}(\bar{z},\rho_0)d\bar{z},
\end{eqnarray}
where $k_\mathcal{R}(\bar{z},\rho_0)$ is given by
\begin{equation}
k_\mathcal{R}(\bar{z},\rho_0)\equiv k(\bar{z},\rho_0)-k_\mathcal{D}(\bar{z},\rho_0).
\end{equation}
We expand $I_\mathcal{R}(\rho_0)$ around $\rho_0=\rho_{\mathrm{sc}}$ as
\begin{equation}
I_\mathcal{R}(\rho_0)=\sum^\infty_{j=0} \frac{1}{j!}(\rho_0-\rho_{\mathrm{sc}})^j \int^1_0 \left. \frac{\partial^j k_\mathcal{R}}{\partial \rho_0^j} \right|_{\rho_0=\rho_{\mathrm{sc}}} d\bar{z}
\end{equation}
and we are only interested in the first term
\begin{equation}
I_\mathcal{R}=\int^1_0 k_\mathcal{R}(\bar{z},\rho_{\mathrm{sc}}) d\bar{z},
\end{equation}
where $k_\mathcal{R}(\bar{z},\rho_{\mathrm{sc}})$ is given by
\begin{equation}
k_\mathcal{R}(\bar{z},\rho_{\mathrm{sc}})= \frac{2}{\bar{z}(2-\bar{z})}\sqrt{\frac{a}{a+2m(1-\bar{z})^3}}- \frac{1}{\bar{z}}\sqrt{\frac{a}{a+2m}}.
\end{equation}
The regular part $I_\mathcal{R}$ is calculated numerically and 
$\bar{b}$ is obtained as
\begin{equation}
\bar{b}= \bar{a} \log \frac{4(a+2m)}{a}+I_\mathcal{R}-\pi.
\end{equation}
Parameters $\bar{a}$ and $\bar{b}$ of the deflection angle~(\ref{eq:al1}) for the images slightly outside of the secondary photon sphere 
in the strong deflection limit $b \rightarrow b_\mathrm{sc}+0 = a+0$ are shown in Fig.~\ref{fig:abcd2}.

\subsection{Case of $a/m>2\sqrt{5}/5$}
\begin{figure}[htbp]
\begin{center}
\includegraphics[width=85mm]{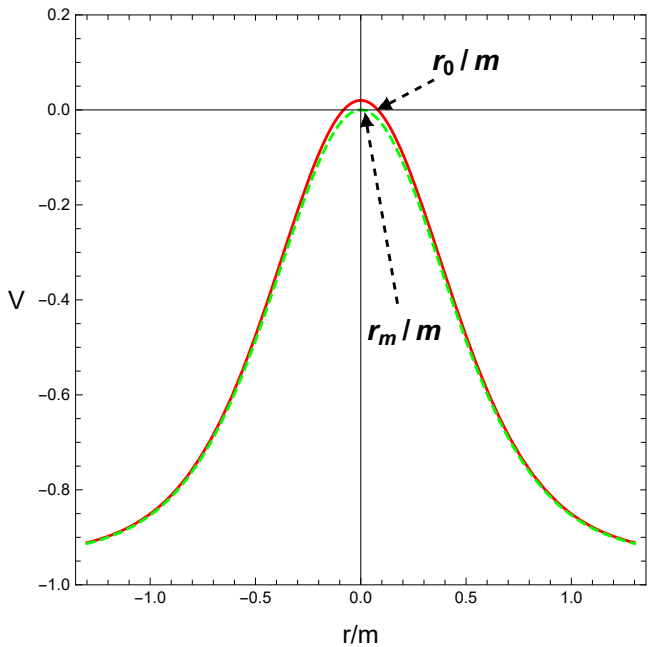}
\end{center}
\caption{Effective potentials~$V$ for images slightly outside of the photon sphere with~$a/m=1$. 
Solid (red) and dashed (green) curves denote the effective potentials~$V$ 
with $b/m=1.01 b_\mathrm{m}/m=1.01$ and with $b/m=b_\mathrm{m}/m=1$, respectively.
The reduced radial coordinates of the photon sphere and the closest distance of the light with $b=1.01 b_\mathrm{m}$ are given by
$r_\mathrm{m}/m=0$ and $r_\mathrm{0}/m=0.0816$, respectively. $E$ is unity.}
\label{fig:V8}
\end{figure}

\begin{figure}[htbp]
\begin{center}
\includegraphics[width=85mm]{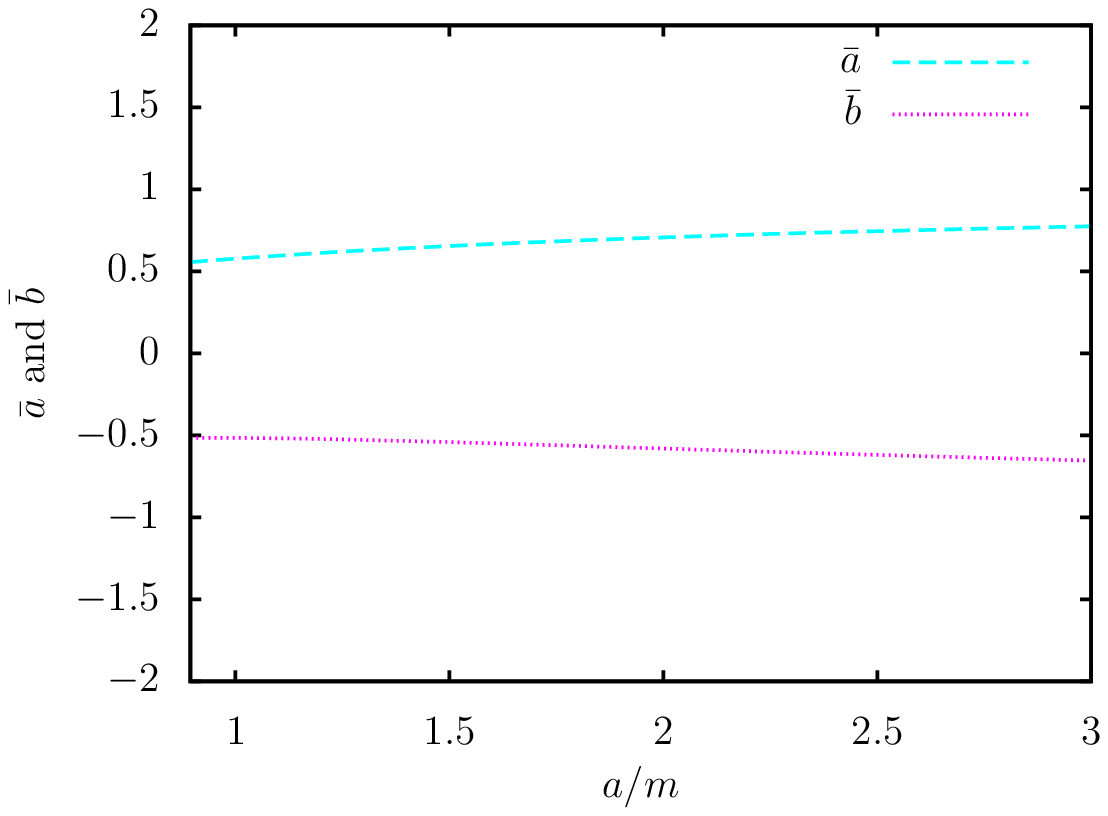}
\end{center}
\caption{Parameters $\bar{a}$ and $\bar{b}$ in the deflection angle~(\ref{eq:al1}) in the strong deflection limit~$b \rightarrow b_{\mathrm{m}}+0$ 
for the images slightly outside of the photon sphere for $a/m>2\sqrt{5}/5$.
Long-dashed cyan and dotted magenta curves denote $\bar{a}$ and $\bar{b}$, respectively.
}
\label{fig:abcd3}
\end{figure}

In the case of $a/m>2\sqrt{5}/5$, the wormhole throat works as the primary photon sphere.
By reading $r_\mathrm{sc}$ in the formulas of Sec.~III-B-3 as $r_\mathrm{m}$, we can use them. 
The effective potential of a light ray to form slightly outside of the primary image is shown in Fig.~\ref{fig:V8}.
Note that $r_{\mathrm{m}}=0$ and $b_{\mathrm{m}}=a$ in this case.
Parameters $\bar{a}$ and $\bar{b}$ in the deflection angle~(\ref{eq:al1}) for the image slightly outside of the photon sphere or the wormhole throat
in the strong deflection limit $r_0\rightarrow r_{\mathrm{m}}+0=+0$ or $b \rightarrow b_{\mathrm{m}}+0=a+0$ are shown in Fig.~\ref{fig:abcd3}.

When $m=0$, 
\begin{equation}
I_\mathcal{D}= -\log \left( \frac{b}{b_{\mathrm{m}}}-1 \right)+ 2\log 2,
\end{equation}
and
\begin{equation}
I_\mathcal{R}=\int^1_0  \frac{1}{2-\bar{z}} d\bar{z} =\log 2.
\end{equation}
Thus, we obtain $\bar{a}=1$ and 
\begin{equation}
\bar{b}=3\log 2 -\pi \sim -1.06
\end{equation}
and we recover the deflection angle by the Ellis-Bronnikov wormhole without the ADM masses in the strong deflection limit 
obtained in Refs.~\cite{Tsukamoto:2016qro,Tsukamoto:2016jzh,Shaikh:2019jfr}.
See Refs.~\cite{Perlick:2003vg,Nandi:2006ds,Tsukamoto:2016zdu,Chetouani_Clement_1984,Tsukamoto:2016qro,Tsukamoto:2016jzh,Shaikh:2019jfr}
for the details of the gravitational lensing in the strong gravitational field.

\section{Gravitational lensing in the strong deflection limit}
\begin{figure}[htbp]
\begin{center}
\includegraphics[width=85mm]{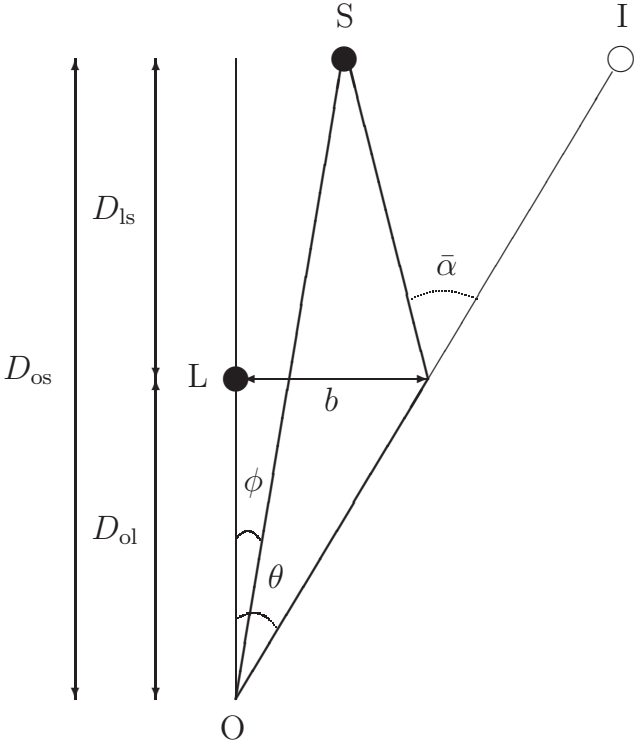}
\end{center}
\caption{Lens configuration. 
A light with an impact parameter $b$ which is emitted by a source S with a source angle $\phi$ and  
which is deflected by a lens L with an effective deflection angle $\bar{\alpha}$, reaches an observer O.
The observer O sees an image I with an image angle $\theta$. 
Distances between the observer O and the source S, between the lens L and the source S, and between the observer O and the lens L, 
are given by $D_{\mathrm{os}}$, $D_{\mathrm{ls}}$, and $D_{\mathrm{ol}}$, respectively.}
\label{fig:configuration}
\end{figure}

We consider a lens configuration shown as Fig.~\ref{fig:configuration}. 
A light ray with an impact parameter $b$ emitted by a source S with a source angle $\phi$ is deflected by a lens L with a deflection angle $\alpha$, 
and an observer O observes the image I of the light ray with an image angle $\theta$. 
We assume that S and O are on the same side of a throat if the lensing object is a wormhole.
We also assume small angles $\bar{\alpha}\ll 1$, $\theta=b/D_{\mathrm{ol}}\ll 1$, and $\phi \ll 1$, 
where $D_{\mathrm{ol}}$ is the distance between the observer O and the lens L,
and $\bar{\alpha}$ is the effective deflection angle of the light ray defined by
\begin{equation}
\bar{\alpha}=\alpha \quad  \mathrm{mod} \quad  2\pi.
\end{equation}
Note that O and S are far away from the photon spheres.
The deflection angle is expressed by 
\begin{equation}\label{eq:defn}
\alpha=\bar{\alpha}+ 2\pi n,
\end{equation}
where $n$ is the winding number of the light ray.

A small-angle lens equation~\cite{Bozza:2008ev} is expressed by
\begin{equation}\label{eq:lens}
D_{\mathrm{ls}}\bar{\alpha}=D_{\mathrm{os}} \left( \theta-\phi \right),
\end{equation}
where $D_{\mathrm{ls}}$ is a distance between the lens L and the source S, 
and $D_{\mathrm{os}}=D_{\mathrm{ol}}+D_{\mathrm{ls}}$ is a distance between the observer O and the source S.
We define $\theta^0_n$ by
\begin{equation}\label{eq:theta0n}
\alpha(\theta^0_n) = 2\pi n,
\end{equation}
and we expand the deflection angle $\alpha(\theta)$ around $\theta=\theta^0_n$ as
\begin{equation}\label{eq:alphaexpand}
\alpha(\theta)=\alpha(\theta^0_n)+\left. \frac{d\alpha}{d\theta} \right|_{\theta=\theta^0_n} (\theta-\theta^0_n)+O\left( (\theta-\theta^0_n)^2 \right).
\end{equation}
We do not consider the primary and marginally unstable photon sphere in the case of $a/m=2\sqrt{5}/5$.

\subsection{Images slightly outside of the primary and secondary photon spheres}
We consider images slightly outside of the primary and secondary photon spheres. 
In this case, we can follow calculations in Ref.~\cite{Bozza:2002zj}. 
The deflection angle $\alpha$ in the strong deflection limit $b\rightarrow b_{\mathrm{m}}+0$ ($b\rightarrow b_{\mathrm{sc}}+0$) is expressed by
\begin{eqnarray}\label{eq:alphaSDL}
\alpha(\theta)
&=&-\bar{a} \log \left( \frac{\theta}{\theta_\infty}-1 \right)+\bar{b} \nonumber\\
&&+O\left( \left( \frac{\theta}{\theta_\infty}-1 \right) \log \left( \frac{\theta}{\theta_\infty}-1 \right) \right),
\end{eqnarray}
where $\theta_\infty \equiv b_{\mathrm{m}}/D_{\mathrm{ol}}$ ($\theta_\infty \equiv b_{\mathrm{sc}}/D_{\mathrm{ol}}$) 
is the image angle of the primary (secondary) photon sphere.
From Eqs.~(\ref{eq:theta0n}) and (\ref{eq:alphaSDL}), $\theta^0_n$ is rewritten by 
\begin{equation}\label{eq:theta0ninf}
\theta^0_n=\left( 1+e^{\frac{\bar{b}-2\pi n}{\bar{a}}} \right) \theta_\infty.
\end{equation}
From 
\begin{equation}
\left. \frac{d\alpha}{d\theta} \right|_{\theta=\theta^0_n}=\frac{\bar{a}}{\theta_\infty-\theta^0_n}
\end{equation}
and Eqs.~(\ref{eq:defn}),  (\ref{eq:theta0n}),  (\ref{eq:alphaexpand}), and (\ref{eq:theta0ninf}),
the effective deflection angle $\bar{\alpha}(\theta_n)$ for $\theta=\theta_n$, which is 
the positive solution of the lens equation with the winding number $n$, is obtained as
\begin{equation}\label{eq:baralpha2}
\bar{\alpha}(\theta_n)=\frac{\bar{a}}{\theta_\infty e^{\frac{\bar{b}-2\pi n}{\bar{a}}}} \left( \theta^0_n-\theta_n \right).
\end{equation}
By substituting Eq.~(\ref{eq:baralpha2}) into the lens equation~(\ref{eq:lens}), we obtain
\begin{equation}
\theta_n(\phi) \sim \theta^0_n - \frac{\theta_\infty e^{\frac{\bar{b}-2\pi n}{\bar{a}}} D_{\mathrm{os}} (\theta^0_n-\phi)}{\bar{a}D_{\mathrm{ls}}}.
\end{equation}
For $\phi=0$, this gives the image angles $\theta_{\mathrm{E}n}$ of relativistic Einstein rings,
\begin{equation}
\theta_{\mathrm{E}n}\equiv \theta_n(0) \sim \left( 1- \frac{\theta_\infty e^{\frac{\bar{b}-2\pi n}{\bar{a}}}D_{\mathrm{os}}}{\bar{a}D_{\mathrm{ls}}} \right) \theta^0_n.
\end{equation}
The difference of image angles between the photon sphere and the outermost image among those slightly outside of the photon sphere is given by
\begin{equation}
\bar{\mathrm{s}}\equiv \theta_1-\theta_\infty \sim \theta^0_1-\theta^0_\infty = \theta_\infty e^{\frac{\bar{b}-2\pi}{\bar{a}}}.
\end{equation}
The magnification of the image angle with the winding number $n$ is obtained as 
\begin{equation}
\mu_n \equiv \frac{\theta_n}{\phi}\frac{d\theta_n}{d\phi} \sim \frac{\theta^2_\infty D_{\mathrm{os}} \left(1+ e^{\frac{\bar{b}-2\pi n}{\bar{a}}}\right) e^{\frac{\bar{b}-2\pi n}{\bar{a}}}}{\phi \bar{a} D_{\mathrm{ls}}}.
\end{equation}
Note that the outermost image with $n=1$ is the brightest one among them.
The sum of the magnifications of all the images and the sum of the magnification of images excluding the outermost image are given by
\begin{eqnarray}
&&\sum^\infty_{n=1} \mu_n \sim \frac{\theta^2_\infty D_{\mathrm{os}} \left(1+e^{\frac{2\pi}{\bar{a}}} +e^{\frac{\bar{b}}{\bar{a}}}\right) e^{\frac{\bar{b}}{\bar{a}}}}{\phi \bar{a} D_{\mathrm{ls}}\left( e^{\frac{4\pi}{\bar{a}}} -1 \right)}, \\
&&\sum^\infty_{n=2} \mu_n \sim \frac{\theta^2_\infty D_{\mathrm{os}} \left(e^{\frac{2\pi}{\bar{a}}}+e^{\frac{4\pi}{\bar{a}}} +e^{\frac{\bar{b}}{\bar{a}}}\right) e^{\frac{\bar{b}-4\pi}{\bar{a}}}}{\phi \bar{a} D_{\mathrm{ls}}\left( e^{\frac{4\pi}{\bar{a}}} -1 \right)}
\end{eqnarray}
and the ratio of the magnifications of the brightest image to the sum of the other images is obtained as
\begin{equation}
\bar{\mathrm{r}} 
\equiv \frac{\mu_1}{\sum^\infty_{n=2} \mu_n}
\sim \frac{\left( e^{\frac{4\pi}{\bar{a}}}-1 \right) \left( e^{\frac{2\pi}{\bar{a}}}+e^{\frac{\bar{b}}{\bar{a}}} \right)}{ e^{\frac{2\pi}{\bar{a}}}+e^{\frac{4\pi}{\bar{a}}} +e^{\frac{\bar{b}}{\bar{a}}} }.
\end{equation}

\subsection{Images slightly inside the primary photon sphere}
We consider the case of the images slightly inside the primary photon sphere as well as  Ref.~\cite{Shaikh:2019itn}.
The deflection angle $\alpha$ in the strong deflection limit $b\rightarrow b_{\mathrm{m}}-0$ is given by
\begin{eqnarray}\label{eq:alphaSDL2}
\alpha(\theta)
&=&-\bar{c} \log \left( 1-\frac{\theta}{\theta_\infty} \right)+\bar{d} \nonumber\\
&&+O\left( \left( 1-\frac{\theta}{\theta_\infty} \right) \log \left( 1-\frac{\theta}{\theta_\infty} \right) \right),
\end{eqnarray}
where $\theta_\infty = b_{\mathrm{m}}/D_{\mathrm{ol}}$ 
is the image angle of the primary photon sphere.
From Eqs.~(\ref{eq:theta0n}) and (\ref{eq:alphaSDL2}), $\theta^0_n$ is rewritten by 
\begin{equation}\label{eq:theta0ninf2}
\theta^0_n=\left( 1-e^{\frac{\bar{d}-2\pi n}{\bar{c}}} \right) \theta_\infty.
\end{equation}
By using 
\begin{equation}
\left. \frac{d\alpha}{d\theta} \right|_{\theta=\theta^0_n}=\frac{\bar{c}}{\theta_\infty-\theta^0_n}
\end{equation}
and Eqs.~(\ref{eq:defn}),  (\ref{eq:theta0n}),  (\ref{eq:alphaexpand}), and (\ref{eq:theta0ninf2}),
we can express the effective deflection angle $\bar{\alpha}(\theta_n)$ as
\begin{equation}\label{eq:baralpha3}
\bar{\alpha}(\theta_n)=\frac{\bar{c}}{\theta_\infty e^{\frac{\bar{d}-2\pi n}{\bar{c}}}} \left( \theta_n-\theta^0_n \right).
\end{equation}
By substituting Eq.~(\ref{eq:baralpha3}) into the lens equation~(\ref{eq:lens}), we obtain the positive position 
\begin{equation}
\theta_n(\phi) \sim \theta^0_n + \frac{\theta_\infty e^{\frac{\bar{d}-2\pi n}{\bar{c}}} D_{\mathrm{os}} (\theta^0_n-\phi)}{\bar{c}D_{\mathrm{ls}}}.
\end{equation}
The image angles of relativistic Einstein rings are given by
\begin{equation}
\theta_{\mathrm{E}n}\sim \left( 1+ \frac{\theta_\infty e^{\frac{\bar{d}-2\pi n}{\bar{c}}}D_{\mathrm{os}}}{\bar{c}D_{\mathrm{ls}}} \right) \theta^0_n.
\end{equation}
The difference of image angles between the photon sphere and the innermost image among those slightly inside the photon sphere is given by
\begin{equation}
\bar{\mathrm{s}}\equiv \theta_1-\theta_\infty \sim \theta^0_1-\theta^0_\infty = -\theta_\infty e^{\frac{\bar{d}-2\pi}{\bar{c}}}.
\end{equation}
The magnification of the image angle given by 
\begin{equation}
\mu_n \sim -\frac{\theta^2_\infty D_{\mathrm{os}} \left(1- e^{\frac{\bar{d}-2\pi n}{\bar{c}}}\right) e^{\frac{\bar{d}-2\pi n}{\bar{c}}}}{\phi \bar{c} D_{\mathrm{ls}}}.
\end{equation}
We notice that the innermost image with $n=1$ is the brightest one among them.
The sum of the magnifications of all the images and the sum of the magnification of images excluding the innermost image are given by
\begin{eqnarray}
&&\sum^\infty_{n=1} \mu_n \sim -\frac{\theta^2_\infty D_{\mathrm{os}} \left(1+e^{\frac{2\pi}{\bar{c}}} -e^{\frac{\bar{d}}{\bar{c}}}\right) e^{\frac{\bar{d}}{\bar{c}}}}{\phi \bar{c} D_{\mathrm{ls}}\left( e^{\frac{4\pi}{\bar{c}}} -1 \right)}, \\
&&\sum^\infty_{n=2} \mu_n \sim -\frac{\theta^2_\infty D_{\mathrm{os}} \left(e^{\frac{2\pi}{\bar{c}}}+e^{\frac{4\pi}{\bar{c}}} -e^{\frac{\bar{d}}{\bar{c}}}\right) e^{\frac{\bar{d}-4\pi}{\bar{c}}}}{\phi \bar{c} D_{\mathrm{ls}}\left( e^{\frac{4\pi}{\bar{c}}} -1 \right)}
\end{eqnarray}
and the ratio of the magnification of the brightest image to the sum of the other images is obtained as
\begin{equation}
\bar{\mathrm{r}} 
\equiv \frac{\mu_1}{\sum^\infty_{n=2} \mu_n}
\sim \frac{\left( e^{\frac{4\pi}{\bar{c}}}-1 \right) \left( e^{\frac{2\pi}{\bar{c}}}-e^{\frac{\bar{d}}{\bar{c}}} \right)}{ e^{\frac{2\pi}{\bar{c}}}+e^{\frac{4\pi}{\bar{c}}} -e^{\frac{\bar{d}}{\bar{c}}} }.
\end{equation}

\section{Gravitational lensing under weak-field approximation}
Under a weak-field approximation $\rho \gg  m$ and $\rho \gg a$, the line element is given, in the usual radial coordinate~$\rho$, by
\begin{eqnarray}
ds^2
&=&-\left( 1-\frac{2m}{\rho} \right)dt^2+\left( 1+\frac{2m}{\rho} \right) \left( 1+\frac{a^2}{\rho^2} \right) d\rho^2 \nonumber\\
&&+\rho^2 (d\vartheta^2+\sin^2\vartheta d\varphi^2)
\end{eqnarray}
which is the same as the Simpson-Visser spacetime with $K=0$ and $N=1$~\cite{Tsukamoto:2020bjm}. 
In this section, we consider both positive and negative impact parameters.

\subsection{$m\neq 0$}
When the ADM mass is nonzero $m\neq 0$, the deflection angle is given by~\cite{Nascimento:2020ime,Ovgun:2020yuv,Tsukamoto:2020bjm}
\begin{eqnarray}\label{eq:weakdef}
\alpha \sim \frac{4m}{b}
\end{eqnarray}
By using Eqs.~(\ref{eq:defn}), (\ref{eq:lens}), and (\ref{eq:weakdef}), and $\theta=b/D_{\mathrm{ol}}$,
the reduced image angles $\hat{\theta}=\hat{\theta}_{\pm 0}$ with the winding number $n=0$ are obtained as
\begin{eqnarray}
\hat{\theta}_{\pm 0} \left( \hat{\phi} \right) =\frac{1}{2} \left( \hat{\phi} \pm \sqrt{\hat{\phi}^2+4} \right),
\end{eqnarray}
where we define $\hat{\theta}\equiv\theta/\theta_{\mathrm{E}0}$ and $\hat{\phi}\equiv\phi/\theta_{\mathrm{E}0}$, 
where $\theta_{\mathrm{E}0}$ is the image angle of the Einstein ring given by
\begin{eqnarray}
\theta_{\mathrm{E}0}\equiv \theta_{+0}(0) \equiv \sqrt{\frac{4mD_{\mathrm{ls}}}{D_{\mathrm{os}}D_{\mathrm{ol}}}}.
\end{eqnarray}
Here and hereinafter, the upper (lower) sign is chosen for the positive (negative) impact parameter.
The magnifications of the images are given by
\begin{eqnarray}
\mu_{\pm 0}
&\equiv& \frac{\hat{\theta}_{\pm 0}}{\hat{\phi}} \frac{d\hat{\theta}_{\pm 0}}{d\hat{\phi}} \nonumber\\
&=& \frac{1}{4} \left( 2\pm \frac{\hat{\phi}}{\sqrt{\hat{\phi}^2+4}} \pm \frac{\sqrt{\hat{\phi}^2+4}}{\hat{\phi}} \right) \nonumber\\
&=& \frac{\hat{\theta}^4_{\pm 0}}{\left(\hat{\theta}^2_{\pm 0} \mp 1 \right)\left(\hat{\theta}^2_{\pm 0} \pm 1 \right)}
\end{eqnarray}
The total magnification of the two images is obtained as
\begin{eqnarray}
\mu_{0\mathrm{tot}}
&\equiv& \left| \mu_{+0} \right|+ \left| \mu_{-0} \right| \nonumber\\
&=& \frac{1}{2} \left( \frac{\hat{\phi}}{\sqrt{\hat{\phi}^2+4}} + \frac{\sqrt{\hat{\phi}^2+4}}{\hat{\phi}} \right). 
\end{eqnarray}

\subsection{$m=0$}
For the vanishing ADM mass $m=0$, the metric corresponds to the Ellis-Bronnikov wormhole
and the deflection angles of light rays are given by~\cite{Abe_2010}
\begin{eqnarray}\label{eq:weakdef2}
\alpha \sim \pm \frac{\pi a^2}{4b^2}.
\end{eqnarray}
By using Eqs.~(\ref{eq:defn}), (\ref{eq:lens}), and (\ref{eq:weakdef2}), and $\theta=b/D_{\mathrm{ol}}$ and 
\begin{eqnarray}
\theta_{\mathrm{E}0}=\left( \frac{\pi a^2 D_{\mathrm{ls}}}{4D_{\mathrm{os}}D^2_{\mathrm{ol}}} \right)^\frac{1}{3},
\end{eqnarray}
the lens equation is expressed by
\begin{eqnarray}
\hat{\theta}^3- \hat{\theta}^2 \hat{\phi}=\pm 1.
\end{eqnarray}
The lens equation has a positive solution $\hat{\theta}=\hat{\theta}_{+ 0}$ and a negative solution $\hat{\theta}=\hat{\theta}_{- 0}$.
The magnifications of the images are given by
\begin{eqnarray}
\mu_{\pm 0}
= \frac{\hat{\theta}^6_{\pm 0}}{\left(\hat{\theta}^3_{\pm 0} \mp 1 \right)\left(\hat{\theta}^3_{\pm 0} \pm 2 \right)}.
\end{eqnarray}
The details of the gravitational lensing by the Ellis-Bronnikov wormhole with vanishing ADM mass
under the weak gravitational field were investigated in Refs.~\cite{Abe_2010}.

\section{Discussion and Conclusion}
A black-bounce spacetime suggested by Lobo~\textit{et al.}~\cite{Lobo:2020ffi}
forms two photon spheres which can be observed by an observer in the parameter region $4\sqrt{3}/9 < a/m \leq  2\sqrt{5}/5$.
We apply the formula in the strong deflection limits to a supermassive black hole candidate at the center of our galaxy. 
As shown in Table~I, the images slightly inside the primary photon sphere are several dozen times brighter than 
the ones slightly outside of the primary photon sphere.
The images slightly outside of the secondary photon sphere are quite fainter than the other images. 
\begin{table*}[htbp]
 \caption{Observables in the strong deflection limits for given $a$ and $m$.
 Parameters~$\bar{a}$, $\bar{b}$, $\bar{c}$, and $\bar{d}$ in the deflection angles~(\ref{eq:al1}) and (\ref{eq:al2}),
 the diameter of the photon sphere~$2\theta_{\infty}$, 
 the diameter of the brightest image~$2\theta_{\mathrm{E}1}$, 
 the difference of the radii of the brightest image and the photon sphere $\bar{\mathrm{s}}=\theta_{\mathrm{E}1}-\theta_\infty$, 
 the magnification of the pair of the brightest images $\mu_{1\mathrm{tot}}(\phi) \sim 2 \left| \mu_{1} \right|$ for the source angle $\phi=1$ arcsecond, 
 and the ratio of the magnification of the brightest image to the other images $\bar{\mathrm{r}}= \mu_1/\sum^\infty_{n=2} \mu_n$
 are shown for the case of $D_{\mathrm{os}}=16$~kpc, $D_{\mathrm{ol}}=D_{\mathrm{ls}}=8$~kpc.
 Here, $m_*$ and $a_*$ are defined as $m_*\equiv 4\times 10^6 M_{\odot}$ and 
 $a_* \equiv 4(2/\pi)^{1/2} (D_{\mathrm{ls}}D_{\mathrm{ol}}/D_{\mathrm{os}})^{1/4} m_*^{3/4}=7.2\times 10^{9}$km, respectively.
 Both the Ellis-Bronnikov wormhole with $m=0$ and $a=a_*$, and the black hole and wormhole spacetimes with $m = m_*$, have 
 the same diameter of Einstein ring, $2\theta_{\mathrm{E}0}=2.86$~arcsecond.
 ISOP, ISOS, and ISIP denote 
 the images slightly outside of the primary photon sphere, 
 the images slightly outside of the secondary photon sphere, and 
 the images slightly inside of the primary photon sphere, respectively.
 Note that $\theta_{\infty}$(ISIP)=$\theta_{\infty}$(ISOP).
}
\begin{center}
\begin{tabular}{ c c c c c c c c c c c c} \hline
$a$           	                   	         &$0$       &$0.5m_*$  &$0.75m_*$  &$0.77m_*$  &$0.8m_*$  &$0.83m_*$  &$0.86m_*$  &$0.9m_*$   	       &$3m_*$  	    &$10m_*$   &$a_*$ \\ 
$m$           	                    	   	 &$m_*$     &$m_*$     &$m_*$      &$m_*$      &$m_*$     &$m_*$      &$m_*$      &$m_*$     	       &$m_*$               &$m_*$     &$0$        \\ \hline
$\bar{a}$(ISOP)        	            	   	 &$1.00$    &$1.08$    &$1.28$     &$1.31$     &$1.39$    &$1.50$     &$1.72$     &$0.56$   	       &$0.78$	            &$0.91$    &$1.00$   \\ 
$\bar{a}$(ISOS)        	            	         &$\cdots$  &$\cdots$  &$\cdots$   &$0.53$     &$0.54$    &$0.54$     &$0.55$     &$\cdots$            &$\cdots$	    &$\cdots$  &$\cdots$   \\ 
$\bar{c}$(ISIP)    	                    	 &$\cdots$  &$\cdots$  &$\cdots$   &$2.6$      &$2.77$    &$3.00$     &$3.43$     &$\cdots$  	       &$\cdots$ 	    &$\cdots$  &$\cdots$ \\ 
$\bar{b}$(ISOP)        	            	   	 &$-0.40$   &$-0.46$   &$-0.73$    &$-0.79$    &$-0.95$   &$-1.22$    &$-1.87$    &$-0.52$ 	       &$-0.65$             &$-0.88$   &$-1.06$ \\ 
$\bar{b}$(ISOS)        	            	   	 &$\cdots$  &$\cdots$  &$\cdots$   &$-0.52$    &$-0.52$   &$-0.52$    &$-0.52$    &$\cdots$            &$\cdots$            &$\cdots$  &$\cdots$ \\ 
$\bar{d}$(ISIP)                           	 &$\cdots$  &$\cdots$  &$\cdots$   &$4.18$     &$3.86$    &$3.30$     &$1.97$     &$\cdots$ 	       &$\cdots$            &$\cdots$  &$\cdots$ \\ 
$2\theta_{\infty}$(ISOP)~[$\mu$as]        	 &$51.58$   &$50.02$   &$47.53$    &$47.23$    &$46.74$   &$46.16$    &$45.48$    &$8.934$     	       &$29.78$             &$99.27$   &$1.203\times10^{4}$ \\ 
$2\theta_{\infty}$(ISOS)~[$\mu$as]        	 &$\cdots$  &$\cdots$  &$\cdots$   &$7.644$    &$7.941$   &$8.239$    &$8.537$    &$\cdots$ 	       &$\cdots$            &$\cdots$  &$\cdots$ \\ 
$2\theta_{\mathrm{E}1}$(ISOP)~[$\mu$as]          &$51.65$   &$50.12$   &$47.73$    &$47.45$    &$46.99$   &$46.47$    &$45.87$    &$8.934$    	       &$29.78$  	    &$99.31$   &$1.204\times10^{4}$ \\ 
$2\theta_{\mathrm{E}1}$(ISOS)~[$\mu$as]          &$\cdots$  &$\cdots$  &$\cdots$   &$7.644$    &$7.941$   &$8.239$    &$8.537$    &$\cdots$    	       &$\cdots$            &$\cdots$  &$\cdots$ \\ 
$2\theta_{\mathrm{E}1}$(ISIP)~[$\mu$as]          &$\cdots$  &$\cdots$  &$\cdots$   &$26.07$    &$27.24$   &$29.08$    &$32.51$    &$\cdots$            &$\cdots$            &$\cdots$  &$\cdots$ \\ 
$\bar{\mathrm{s}}$(ISOP)~[$\mu$as]               &$0.032$   &$0.048$   &$0.098$    &$0.11$     &$0.13$    &$0.16$     &$0.20$     &$2.2\times10^{-5}$  &$1.9\times10^{-3}$  &$0.020$   &$3.9$ \\ 
$\bar{\mathrm{s}}\times10^{6}$(ISOS)~[$\mu$as]   &$\cdots$  &$\cdots$  &$\cdots$   &$9.5$      &$12$      &$14$       &$18$       &$\cdots$            &$\cdots$            &$\cdots$  &$\cdots$ \\ 
$\bar{\mathrm{s}}$(ISIP)~[$\mu$as]               &$\cdots$  &$\cdots$  &$\cdots$   &$-10.6$    &$-9.75$   &$-8.54$    &$-6.49$    &$\cdots$            &$\cdots$            &$\cdots$  &$\cdots$ \\ 
$\mu_{1\mathrm{tot}}(\phi)\times10^{17}$(ISOP)   &$1.6$     &$2.2$     &$3.6$      &$3.8$      &$4.2$     &$4.7$      &$5.1$      &$3.5\times10^{-4}$  &$0.072$             &$2.1$     &$4.5\times10^{4}$ \\ 
$\mu_{1\mathrm{tot}}(\phi)\times10^{21}$(ISOS)   &$\cdots$  &$\cdots$  &$\cdots$   &$1.3$      &$1.7$     &$2.1$      &$2.7$      &$\cdots$            &$\cdots$            &$\cdots$  &$\cdots$ \\ 
$\mu_{1\mathrm{tot}}(\phi)\times10^{17}$(ISIP)   &$\cdots$  &$\cdots$  &$\cdots$   &$102$      &$93$      &$80$       &$60$       &$\cdots$            &$\cdots$	    &$\cdots$  &$\cdots$ \\ 
$\bar{\mathrm{r}}$(ISOP)     	                 &$535$     &$344$     &$137$      &$119$      &$92.2$    &$65.2$     &$38.1$     &$7.91\times10^{4}$  &$3.33\times10^{3}$  &$975$     &$535$ \\ 
$\bar{\mathrm{r}}\times10^{-4}$(ISOS)    	 &$\cdots$  &$\cdots$  &$\cdots$   &$15.0$     &$12.7$    &$10.9$     &$9.47$     &$\cdots$            &$\cdots$	    &$\cdots$  &$\cdots$ \\ 
$\bar{\mathrm{r}}$(ISIP)     	                 &$\cdots$  &$\cdots$  &$\cdots$   &$5.70$     &$5.24$    &$4.67$     &$3.89$     &$\cdots$            &$\cdots$	    &$\cdots$  &$\cdots$ \\ 
\hline
\end{tabular}
\end{center}
\end{table*}

For simplicity, we concentrate on only positive impact parameters or image angles in the strong deflection limits. 
Notice that the lens equation~(\ref{eq:lens}) with the winding number $n\geq 1$ 
has a negative solution $\theta \sim -\theta_n$ for each $n\geq 1$
and it makes a pair together with the positive image angle $\theta_n$. 
The diameter of the pair of the images is given by 
$2 \theta_n$.
The magnification of the negative image angle is given by $-\mu_{n}$ approximately.
The total magnification of the pair of images is obtained as $\mu_{n \mathrm{tot}} \sim 2 \left| \mu_{n} \right|$.

\begin{figure}[htbp]
\begin{center}
\includegraphics[width=85mm]{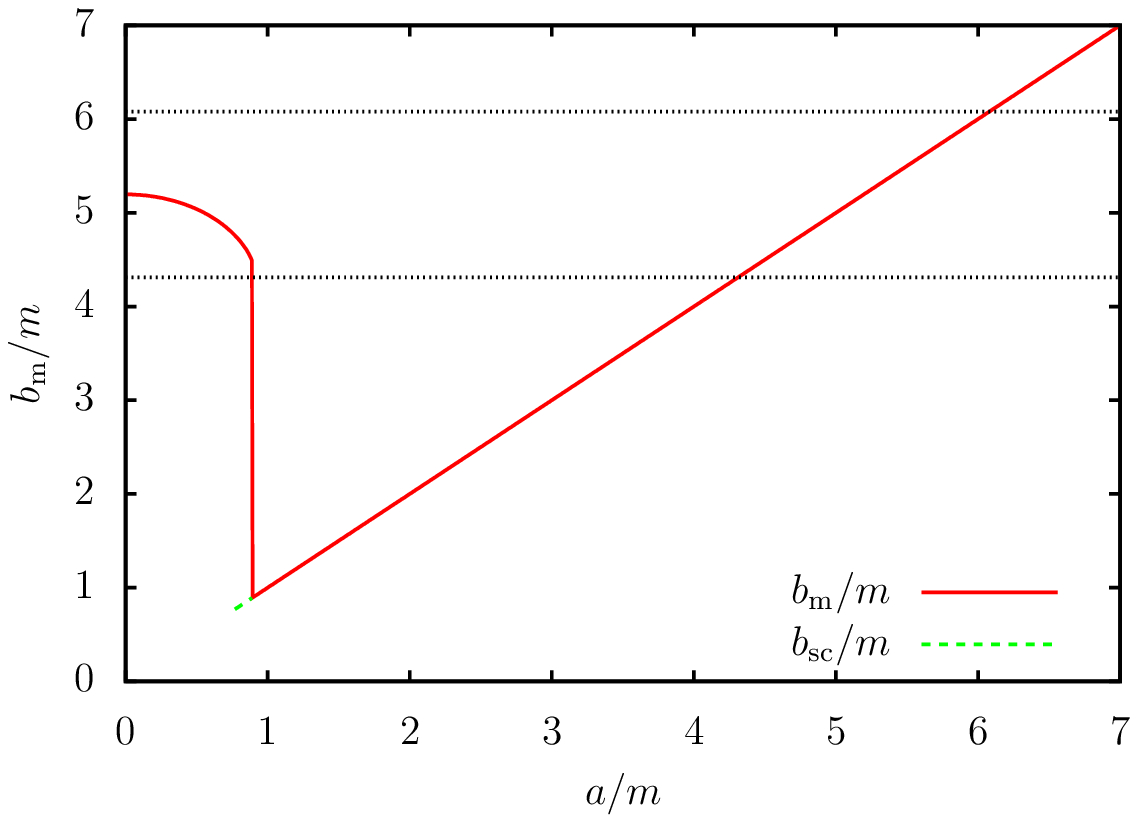}
\end{center}
\caption{The impact parameters for photon spheres.
Solid (red) curve denotes the specific impact parameter $b_\mathrm{m}/m$ of the primary photon sphere as a function of $a/m$.
The specific impact parameter $b_\mathrm{sc}/m$ of the secondary photon sphere is also shown as a dashed (green) line.
Dotted lines show the upper and the lower bounds $4.31<b_\mathrm{m}/m<6.08$ for the observed shadow at the center of M87.}
\label{fig:bm}
\end{figure}

We also note that the observation of the center of galaxy M87 by the Event Horizon Telescope Collaboration~\cite{Akiyama:2019cqa,Akiyama:2019eap,Kocherlakota:2021dcv} 
implies that a large deviation from the Schwarzschild metric and Kerr metric is excluded.
If we assume that the observed shadow size is the same as the size of the primary photon sphere,
the critical impact parameter must be within $4.31<b_\mathrm{m}/m<6.08$ at $68$\% confidence levels.
Figure~\ref{fig:bm} shows 
that a Schwarzschild black hole for $0 = a/m$, 
a regular black hole for $0 < a/m<4\sqrt{3}/9$, 
a wormhole with two photon spheres for $4\sqrt{3}/9 \leq a/m<2\sqrt{5}/5$, 
and a wormhole with a photon sphere for $4.31<a/m<6.08$ could survive.
However, the effects of the throat on the shadow may have subtle problems, and we would need simulations to treat them.

As shown in this paper, in the spacetime suggested by Lobo~\textit{et al.}~\cite{Lobo:2020ffi}, 
the wormhole throat acts as a photon sphere when another photon sphere is outside of the throat. 
This is not a general property for static and spherical symmetric wormholes with reflection symmetry.
The throats of a Damour-Solodukhin wormhole~\cite{Damour:2007ap} and a wormhole in the Simpson-Visser spacetime~\cite{Simpson:2018tsi}
act as an antiphoton sphere when there is a photon sphere outside of the throat.
Thus, the shadow images and gravitational lensing by multiple photon spheres in strong deflection limits 
are not universal properties for reflection-symmetric wormholes.
Moreover, while we have focused on the photon spheres, 
we should remember that shadow and light ring images can be formed in spacetimes without photon spheres~\cite{Chiba:2017nml,Paul:2020ufc,Dey:2020bgo}.

We mention that our work does not calculate all the lens configurations in the spacetime.
If a light source is on a different side of the wormhole throat, 
images are formed inside the throat. 
In this case, one may use an exact lens equation investigated by Perlick~\cite{Perlick:2003vg,Tsukamoto:2016zdu},
or one may use an approximate lens equation with the deflection angle in a strong deflection limit~\cite{Shaikh:2019jfr}.
However, the errors caused by the approximation under the lens configuration has not been studied.  

We do not consider the gravitational lensing by a marginally unstable photon sphere for $a/m=2\sqrt{5}/5$.
In this case, we cannot use Eqs.~(\ref{eq:al1}) and (\ref{eq:al2}) since the deflection angles do not logarithmically diverge in the strong deflection limit.
We can use formulas in Ref.~\cite{Tsukamoto:2020iez} for the lensed images slightly outside of the marginally unstable photon sphere, 
but we cannot use them for images slightly inside the marginally unstable photon sphere.
The latter case is left as one for future works.

%
\appendix
\section{Variable $z$ and its alternatives}
Bozza~\cite{Bozza:2002zj} introduced a variable to formalize the deflection angle of light in a strong deflection limit $b \rightarrow b_\mathrm{m}+0$
in a general static and spherical spacetime, in the signature $(+,-,-,-)$:
\begin{equation}
z_{[28](+,-,-,-)}\equiv \frac{g_{tt}(r)-g_{tt}(r_0)}{1-g_{tt}(r_0)},
\end{equation}
and which is correspondent with the following, in the signature $(-,+,+,+)$:
\begin{equation}\label{eq:zBozza}
z_{[28](-,+,+,+)}= \frac{-g_{tt}(r)+g_{tt}(r_0)}{1+g_{tt}(r_0)}.
\end{equation}
For the line element~(\ref{eq:metric}), we obtain
\begin{equation}\label{eq:zBozza2}
z_{[28](-,+,+,+)}=\frac{A-A_0}{1-A_0}=1- \left( \frac{r_0^{2N}+a^{2N}}{r^{2N}+a^{2N}} \right)^\frac{K+1}{2N} \left( \frac{r}{r_0} \right)^K.
\end{equation}
The derivative of $z_{[28](-,+,+,+)}$ with respect to $r$ is given by
\begin{equation}
\frac{dz_{[28](-,+,+,+)}}{dr}=-\frac{\left(r_0^{2N}+a^{2N}\right)^\frac{K+1}{2N}}{r_0^K} \frac{r^{K-1}\left(Ka^{2N}-r^{2N} \right)} {\left( r^{2N}+a^{2N} \right)^{\frac{K+1}{2N}+1}}.
\end{equation}
The variable $z_{[28](-,+,+,+)}$ for $K\neq 0$ is not suitable for a strong deflection limit analysis, 
since it is not a monotonic variable when $r$ runs across $r=K^{\frac{1}{2N}}a$.
Thus, we cannot apply Bozza's formula directly for the spacetime with $K\neq 0$ in the Buchdahl coordinates. 

Nascimento~\textit{et al.}~\cite{Nascimento:2020ime} have investigated gravitational lensing in the strong deflection limit in the Simpson-Visser spacetime 
in a signature $(-,+,+,+)$ with $K=0$, $N=1$, and $m>3a$ by using an alternative variable 
\begin{equation}\label{eq:zT}
z\equiv 1-\frac{r_0}{r}
\end{equation}
which is suggested by Tsukamoto in Ref.~\cite{Tsukamoto:2016jzh}. 
This variable works in the Simpson-Visser spacetime, as shown in Ref.~\cite{Nascimento:2020ime}.

In Ref.~\cite{Tsukamoto:2020bjm}, 
Tsukamoto has investigated the strong deflection limit analysis in the Simpson-Visser spacetime in a signature $(-,+,+,+)$ 
with $K=0$, $N=1$, and any non-negative parameters $m$ and $a$. 
We notice that the definition of the variable in Ref.~\cite{Tsukamoto:2020bjm} has a typo which does not affect the other equations. 
We should read Eq.~(3.1) in Ref.~\cite{Tsukamoto:2020bjm} 
\begin{equation}\label{eq:zTsukamoto1}
z\equiv \frac{g_{tt}(r)-g_{tt}(r_0)}{1-g_{tt}(r_0)}=1-\frac{\sqrt{r_0^2+a^2}}{\sqrt{r^2+a^2}}
\end{equation}
as 
\begin{equation}\label{eq:zTsukamoto2}
z\equiv \frac{-g_{tt}(r)+g_{tt}(r_0)}{1+g_{tt}(r_0)}=1-\frac{\sqrt{r_0^2+a^2}}{\sqrt{r^2+a^2}}.
\end{equation}
By inserting $K=0$ and $N=1$ into Eq.~(\ref{eq:zBozza2}), we can recover Eq.~(\ref{eq:zTsukamoto2}).
In the Simpson-Visser spacetime, the variable~(\ref{eq:zTsukamoto2}) also works well 
in the strong deflection limit analysis, as shown in Ref.~\cite{Tsukamoto:2020bjm}.
Notice that the variable $z$ (\ref{eq:zTsukamoto2}) in Ref.~\cite{Tsukamoto:2020bjm} 
corresponds with $\bar{z}$~(\ref{eq:barz}) in this paper.

\end{document}